%% file: HNN_2005.tex
\documentclass[draft]{elsart}%
\usepackage[english]{babel}%
\usepackage{amssymb,amsmath,graphicx}%
\def\Elproofname{Proof.}
\input{macros}
\begin{document}%
\begin{frontmatter}


\title{On a thermodynamically consistent modification of the Becker-D\"{o}ring equations}

\author[HUB,MH]{M. Herrmann},
\author[HUB,DFG]{M. Naldzhieva},
\author[HUB,DFG,BN]{B. Niethammer}
\address[HUB]{%
Humboldt Universit\"at zu Berlin, Department of Mathematics,\\
Section for Applied Analysis,
\\
Unter den Linden 6,
D-10099 Berlin,
Germany%
}%

\thanks[MH]{E-Mail {\tt{michaelherrmann@mathematik.hu-berlin.de}}}
\thanks[DFG]{supported by the DFG research center \textsc{Matheon}, see \texttt{www.matheon.de}}
\thanks[BN]{E-Mail {\tt{niethamm@mathematik.hu-berlin.de}}}

\begin{abstract}
Recently, Dreyer and Duderstadt have proposed a modification of the
Becker--D\"{o}ring cluster equations which now have a nonconvex
Lyapunov function. We start with existence and uniqueness results
for the modified equations. Next we derive an explicit criterion for
the existence of equilibrium states and solve the minimization
problem for the Lyapunov function. Finally, we discuss the long time
behavior in the case that equilibrium solutions do exist.
\end{abstract}
\begin{keyword}
Becker--D\"{o}ring equations 
\sep 
coagulation and fragmentation 
\sep
nonconvex Lyapunov function
\sep 
existence of equilibrium 
\sep  
convergence to equilibrium
\PACS 05.45 \sep 86.03 \sep 82.60
\end{keyword}
\end{frontmatter}
\newcommand{\mhinclude}[1]{\input{#1}}
\input{ch_intro}%
\input{ch_exist}
\input{ch_equi}
\input{ch_lim}
\begin{ack}
The authors would like to thank Wolfgang Dreyer and Frank Duderstadt for several
fruitful discussions and for their valuable comments and remarks.
\end{ack}
\bibliographystyle{elsart-num}
\bibliography{bede}
\end{document}

%% file: macros.tex
\newcommand{\tdots}{{...}}
\newcommand{\eps}{\varepsilon}
\newcommand{\rra}{\rightarrow}
\newcommand{\discr}[2]{#1^{\at{#2}}}

\newcommand{\equi}[1]{\overline{#1}}

\newcommand{\vap}{\mathrm{V}}
\newcommand{\cond}{\mathrm{C}}

\newcommand{\const}{\mathrm{const}}

\newcommand{\ND}{N}
\newcommand{\NL}{N}
\newcommand{\GammaCondL}{\gamma_{\,l}}
\newcommand{\ie}{i.e.}
\newcommand{\D}{\displaystyle}

\newcommand{\ellOne}{\ell^1(\mathbb{N})}
\newcommand{\ellInf}{\ell^\infty(\mathbb{N})}
\newcommand{\pair}[2]{{\left({#1},\,{#2}\right)}}

\newcommand{\at}[1]{{\left({#1}\right)}}
\newcommand{\bat}[1]{{\big(#1\big)}}
\newcommand{\Bat}[1]{{\Big(#1\Big)}}

\newcommand{\ocinterval}[2]{(#1,\,#2]}%
\newcommand{\cointerval}[2]{[#1,\,#2)}%
\newcommand{\oointerval}[2]{(#1,\,#2)}%
\newcommand{\ccinterval}[2]{[#1,\,#2]}%
\newcommand{\norm}[1]{\left|\!\left|{#1}\right|\!\right|}

\newcommand{\abs}[1]{\left|{#1}\right|}

%% file: ch_intro.tex
\section{Introduction}\label{secIntro}%
The Becker--D\"{o}ring equations are an infinite set of kinetic
equations that describe the dynamics of cluster formation in a
system of identical particles. In this model, clusters can coagulate
to form larger clusters or fragment to smaller ones. In what follows
we describe clusters by their size $l\geq2$, the number of particles
in the cluster, and we denote by $z_l\at{t}$ the total number of
$l$--clusters in the system at time $t$. Note that here we always
assume that all $l$-clusters are uniformly distributed in the
physical space. Moreover, the number of free atoms in the system is
abbreviated with $z_1\at{t}$, so that the state of the complete
system is given by a nonnegative sequence
$z\at{t}=\at{z_l\at{t}}_{l\in\Nset}$, where $0\not\in\Nset$.
\par%
The crucial assumption of Becker and D\"{o}ring in \cite{BD35} was
that an $l$--cluster can change its size only by gaining a free atom
(\emph{coagulation}) to form an $(l+1)$--cluster, or loosing an atom
(\emph{fragmentation}) to form an $(l-1)$--cluster. In particular,
for all $l\geq2$ there are two typical transition rates, namely a
\emph{condensation rate} $\Gamma^\cond_l\at{t}$ and a
\emph{vaporization rate} $\Gamma^\vap_l\at{t}$ giving at time $t$
the probability that a $l$-clusters gains or looses a
$1$-cluster, respectively. The net rate of conversion of
$l$--clusters into $(l+1)$--clusters is denoted by $J_l\at{t}$. For
$l\geq{2}$ it reads
\begin{align}
\label{BD1} \tag{BD1}
J_l\at{t}&=\Gamma_l^\cond\at{t}\,{z_l}\at{t}-%
\Gamma_{l+1}^\vap\at{t}\,{z_{l+1}\at{t}},
\end{align}
and the change of the total number of $l$-clusters for $l\geq{2}$ is
given by
\begin{align}
\label{BD2} \tag{BD2}%
\frac{\mathrm{d}}{\mathrm{d}t}z_l\at{t}&=%
J_{l-1}\at{t}-J_l\at{t},\quad{l\geq{2}}.
\end{align}
To describe the change of $z_1\at{t}$, the number of free atoms, a
different equation is needed because free particles are involved in
all reactions in the system. Here we are only interested in the case
that the total number of all atoms in the system is conserved, {\ie}
$\varrho\at{z\at{t}}=\const$, where
\begin{align}
\label{IntroMassDef}%
\varrho\at{z}&=\sum\limits_{l=1}^{\infty}lz_l.
\end{align}
This constraint gives rise to
\begin{math}
\frac{\d}{\d{t}}{z}_1\at{t}=-J_1\at{t} -\sum_{l=1}^{\infty}J_l\at{t},
\end{math}
which can be expressed as follows
\begin{align}
\tag{BD3} \label{BD3} %
\frac{\d}{\d{t}}{z}_1\at{t}=J_{0}\at{t}-J_1\at{t},\quad
J_0\at{t}=-\sum\limits_{l=1}^{\infty}J_l\at{t}.
\end{align}
The system \eqref{BD1}--\eqref{BD3} was derived and investigated
the first time by Frenkel in \cite{Fre39}. Clearly, the equations must be
closed by some \emph{constitutive} \emph{assumptions} relating the
rates $\Gamma^\cond_l\at{t}$ and $\Gamma^\vap_l\at{t}$ to the state
$z\at{t}$ of the system.
\par%
In \cite{DD05}, Dreyer and Duderstadt give a historical overview on
the Becker--D\"{o}ring equations with mass
conservation. As they point out, almost all of the literature is based
on a misinterpretation of \cite{BD35}: The quantities $z_l\at{t}$
are considered as the \emph{volume densities} of $l$-clusters, and
not as \emph{numbers}. Clearly, this reinterpretation corresponds to
the non-explicit assumption that the total volume of the system is
conserved. Dreyer and Duderstadt criticize this
standard interpretation and the resulting constitutive laws, and
derive new closure laws from fundamental thermodynamic
principles. Next we first summarize the standard model, and
afterwards we describe the modified model in detail.
\subsubsection*{The standard model}%
In the standard model, see for instance
\cite{Bur77,PL79,Pen89}, the dynamical equations
\eqref{BD1}--\eqref{BD3} are closed by the following constitutive
assumptions
\begin{align}
\label{SM}\tag{SM}%
\Gamma_l^\cond\at{t}=c_l\,z_1\at{t},\quad
\Gamma_l^\vap\at{t}=d_{\,l},
\end{align}
where $c_l$ and $d_l$ depend neither on the state $z$ nor on the
time $t$. In fact, this is reasonable if $z_1\at{t}$ is the volume
density of free atoms. The coefficient $c_l$ and $d_l$ are then
determined by some heuristic arguments. To give an example, a
very common ansatz is
\begin{align}
\label{SM2}%
c_l =  l^{\alpha},\quad
d_{\,l}=c_l \Big (z_s + \frac{q}{l^\gamma} \Big )%
\end{align}
 with $0\leq\alpha<1$, $z_s>0$, $q>0$, $\gamma<1$, and
\begin{align*}
\begin{array}{lll}
 \alpha=1/3,&\quad\gamma=1/3&\quad\quad\quad\quad%
 \text{for \emph{diffusion controlled kinetics} in 3D},\\
 \alpha=0,&\quad\gamma = 1/2&\quad\quad\quad\quad%
 \text{for \emph{diffusion controlled kinetics} in 2D},\\
 \alpha=2/3,&\quad\gamma=1/3&\quad\quad\quad\quad%
 \text{for \emph{interface reaction limited kinetics} in 3D},\\
 \alpha = 1/2,&\quad\gamma=1/2&\quad\quad\quad\quad%
 \text{for \emph{interface reaction limited kinetics} in 2D}.
\end{array}
\end{align*}

Within the standard model \eqref{BD1}--\eqref{BD3} with \eqref{SM}
there exists a convex Lyapunov function $L$ with
\begin{align}
\label{FreeEnergyClassModel}
L\at{z}&=\sum\limits_{l=1}^{\infty}%
z_l\at{\ln\at{\frac{z_l}{Q_l}}-1},\quad
Q_1=1,\quad{}Q_{l+1}=\prod\limits_{n=1}^{l}\frac{c_{n}}{d_{n+1}},
\end{align}
such that $L\at{z\at{t}}$ decreases with time $t$ for all solutions
$z\at{t}$. An equilibrium state $\equi{z}$ of the dynamics is a state for
which all transfer rates $J_l$ vanish. After some basic calculation
we find that an equilibrium state $\equi{z}$ and its
density $\equi\varrho$ are given by
\begin{align}
\label{IntroStandardEqn3}%
\equi{z}_l=Q_l\,\mu^l,\quad%
\equi{\varrho}=\sum\limits_{l=1}^{\infty}\,l\,Q_l\,\mu^l.
\end{align}
With \eqref{SM2} it can be shown that the radius of convergence of the
power series in \eqref{IntroStandardEqn3} is $z_s$, and that for
$\mu=z_s$ the series converges to
$\varrho_s=\sum_{l=1}^{\infty}lQ_lz_s^l$. In particular,
$\varrho_s$ is the maximal value for the equilibrium density,
and can be interpreted as saturation density. As a
consequence, if the density $\varrho_0$ of initial data exceeds
$\varrho_s$, for $t\rra\infty$ the total mass of the system
cannot be stored in a equilibrium solution, but the excess density
$\varrho_0-\varrho_s$ must be transferred into larger and
larger clusters when time proceeds. However, this process 
is in general extremly slow if the excess density is small. 
This metastability has been rigorously established in \cite{Pen89} 
for typical initial data. As a consequence, exact numerical 
simulations are difficult to perform and impossible to perform 
for small $\varrho_0-\varrho_s$, see \cite{CDW95}. In addition, 
it has been established that the dynamics of large clusters 
after the metastable state can be described the classical 
Lifshitz-Slyozov-Wagner equation for coarsening \cite{Pen89,Sle00,Nie03}.
\subsubsection*{The non-standard model of Dreyer and Duderstadt}%
Dreyer and Duderstadt \cite{DD05} model the system of all clusters=\emph{droplets} 
as mixture of different substances, where a droplet with $l$ atoms is regarded as
a particle of the substance $l$. To be more precise, Dreyer and
Duderstadt introduce a maximal size $l_{\max{}}$ for the droplets,
and thus they consider a mixture of $l_{\max{}}$ different
substances. Since the maximal droplet size $l_{\max{}}$ is usually
very large, we are mainly interested in the limiting case
$l_{\max{}}=\infty$.
\par%
The main advantage of this new approach is that thermodynamics is
able to describe the equilibrium without any knowledge of the
dynamical law. On the contrary, thermodynamics give some constraints
for the dynamical law. The main ideas in \cite{DD05} can be
summarized as follows.
\begin{enumerate}
\item
The Second Law of thermodynamics states that the \textit{available
free energy}, or \emph{availability}, of the system becomes minimal
in equilibrium. This follows from a careful evaluation of Clausius
theorem, and reflects the assumption on the physical process.
\item The available free energy $a_l$ for a single droplet with $l$
atoms can be given explicitly in many situations, see for instance
the examples below.
\item Thermodynamic mixture theory provides an explicit expression
for the availability $A$ of a many droplet system. In particular,
with $a_1=0$ it follows that
\begin{align} \label{FreeEnergyNewModel1}
A\at{z}&=\sum\limits_{l=1}^{\infty}a_l\,z_l+
\sum\limits_{l=1}^{\infty}z_l\ln\at{\frac{z_l}{\ND\at{z}}},
\end{align}
where $\ND\at{z}$ abbreviates the total number of all droplets,
{\ie}
\begin{align}
\ND\at{z}&=\sum\limits_{l=1}^{\infty}z_l.
\end{align}
Note that the second sum in \eqref{FreeEnergyNewModel1} takes care
of the \emph{entropy of mixing}.
\item
The Second Law of thermodynamics requires that the availability
$A$ decreases with time for any real world process, and from this
we obtain a consistency relation for the transition rates
$\Gamma_l^\vap$ and $\Gamma_l^\cond$, see below.
\end{enumerate}
For convenience we set
\begin{align}
\label{FreeEnergyNewModel2}
q_l&=\exp\at{-a_l}\quad\text{with}\quad{q_1}=1,
\end{align}
so that the availability $A$ of the many-droplet system reads
\begin{align}
\label{FreeEnergyNewModel3} A\at{z}=\sum\limits_{l=1}^{\infty}z_l
\ln\at{\frac{z_l}{q_l\ND\at{z}}}= %
\sum\limits_{l=1}^{\infty}z_l\,\ln{z_l}-%
\sum\limits_{l=1}^{\infty}z_l\,\ln{q_l}-%
\ND\at{z}\ln{\ND\at{z}}.
\end{align}
Since the function $x\mapsto{x}\ln{x}$ is convex, we conclude that $A$ is the 
sum of a convex, a linear and a concave functional. 
In particular, $A$ is a neither convex nor concave.
\par%
Next we evaluate the thermodynamic consistency relation mentioned
above. A formal calculation yields
\begin{align}
\notag%
\frac{\d}{\d{t}}{A}\at{z}&=%
\at{\sum\limits_{l=1}^{\infty}\at{1+\ln\frac{z_l}{q_l}}{\frac{\d}{\d{t}}{z}_l}}
-\Bat{1+\ln{\ND\at{z}}}\frac{\d}{\d{t}}{\ND}\at{z}%
\\&=\notag%
\sum\limits_{l=1}^{\infty}%
\ln\at{\frac{z_l}{q_l\,\ND\at{z}}}\frac{\d}{\d{t}}{z}_l
=\notag%
\sum\limits_{l=1}^{\infty}%
\Bat{J_{l-1}\at{z}-J_l\at{z}}\,%
\ln\at{\frac{z_l}{q_l\,\ND\at{z}}}
\\&=\notag%
\ln\at{\frac{z_1}{\ND\at{z}}}J_0\at{z}+
\sum\limits_{l=1}^{\infty}%
J_l\at{z}\,%
\at{\ln\at{\frac{z_{l+1}}{q_{l+1}\,\ND\at{z}}}-
\ln\at{\frac{z_l}{q_l\,\ND\at{z}}}}
\\&=\notag%
\ln\at{\frac{z_1}{\ND\at{z}}}%
\at{-\sum\limits_{l=1}^{\infty}J_l\at{z}}+
\sum\limits_{l=1}^{\infty}%
J_l\at{z}\,%
\ln\at{\frac{q_l\,z_{l+1}}{q_{l+1}\,z_{l}}}
\\&=\notag%
\sum\limits_{l=1}^{\infty}%
J_l\at{z}\,%
\ln\at{\frac{q_l\,z_{l+1}\,\ND\at{z}}{q_{l+1}\,z_{l}\,z_1}}
\\&=\label{FreeEnergyNewModel4}%
\sum\limits_{l=1}^{\infty}%
\at{\Gamma_l^\cond{z_l}-\Gamma_{l+1}^\vap{z_{l+1}}}\,%
\ln\at{\frac{q_l\,z_{l+1}\,\ND\at{z}}{q_{l+1}\,z_{l}\,z_1}}
\end{align}
The Second Law of thermodynamics states that $\frac{\d}{\d{t}}{A}\at{z}$ is
non-positive for all solutions of the Becker-D\"{o}ring dynamics
\eqref{BD1}--\eqref{BD3}. Dreyer and Duderstadt satisfy this restriction
by fixing the ratio between the transition rates via
\begin{align}
\label{NSM}\tag{NSM}%
\frac{\Gamma_{l+1}^\vap\at{t}}{\Gamma_{l}^\cond\at{t}}&=
\frac{q_l}{q_{l+1}}\frac{\ND\at{z\at{t}}}{z_1\at{t}},
\end{align}
so that the net rates $J_l\at{t}$ for $l\geq{1}$ read
\begin{align}
J_l\at{t}={\Gamma_{l}^\cond}\at{t}\,
\at{z_l\at{t}-\frac{\ND\bat{z\at{t}}}{z_1\at{t}}\,%
\frac{q_l\,{z_{l+1}\at{t}}}{q_{l+1}}}.%
\end{align}
With \eqref{NSM} the production of availability becomes
\begin{align}
\notag%
\frac{\d}{\d{t}}{A}\at{z}&=%
\sum\limits_{l=1}^{\infty}\,{\Gamma_{l}^\cond}\,%
\at{z_l-\frac{\ND\at{z}}{z_1}\frac{q_l}{q_{l+1}}{z_{l+1}}}
\ln\at{\frac{q_l\,z_{l+1}\,\ND\at{z}}{q_{l+1}\,z_{l}\,z_1}}%
\\&=%
\sum\limits_{l=1}^{\infty}\,{\Gamma_{l}^\cond}\,%
\bat{z_l-w_l}\,\bat{\ln{w_l}-\ln{z_l}}\leq0,%
\end{align}
with $w_l=\ND\at{z}\,z_{l+1}\,q_l/\at{q_{l+1}\,z_1}$. In particular,
the availability $A$ is a nonconvex Lyapunov function for the
dynamical system \eqref{BD1}--\eqref{BD3} with \eqref{NSM}. 
\par%
In \cite{DD05}, Dreyer and Duderstadt derive the availability $A$
for two important examples. As mentioned above, they always consider
a system which contains only a single droplet with $l$ atoms, and
derive explicit expression for the availability $a_l$. The
availability  $A$ of the many-droplet system is given by
\eqref{FreeEnergyNewModel1}.
\par\emph{Example 1} %
corresponds to a simple vapor-liquid system, in which a single
gaseous droplet with $l$ atoms is included in a liquid matrix, both
made from the same chemical substance as for instance water. The
result is
\begin{align}
\label{BeDeEx1Eqn1} %
a_0=1,\quad{}a_l&=-\delta{l}+\gamma{l}^{\frac{2}{3}}\quad%
\text{for $l>1$}.%
\end{align}
where $\delta$ and $\gamma$ are positive constants.
\par\emph{Example 2} %
is more complicated, and describes a single liquid droplet contained
in a crystalline solid, where both are a binary mixture of Gallium
and Arsenic. Moreover, the solid is surrounded by an inert gas with
prescribed pressure. The resulting expressions for the availability
show that $a_l$ growth with $l$ for large $l$, and this gives rise
to the following simplified ansatz
\begin{align}
\label{BeDeEx2Eqn1} %
a_l&=+\beta\,l\quad\text{for }l\gg1,
\end{align}
where $\beta$ is a positive constant.
\noindent%
We will show in Section \ref{secEqui} that both examples differ in
the set of possible equilibrium states.
\par%
Although thermodynamics give a constraint for the dynamical law, we
are free to choose the transition rates $\Gamma^\cond_l\at{t}$. In
what follows we always assume that
\begin{align}
\label{ConstLawFlux2} %
\Gamma^\cond_l\at{t}=z_1\at{t}\,\GammaCondL,
\end{align}
where $\GammaCondL$ is constant. We mention that other choices of
the time dependence of $\Gamma^\cond_l\at{t}$ may be reasonable, which,
however, change only the time scale of the evolution. Finally, we
obtain the following system of equations
\begin{align}
\label{MBD1}\tag{MBD1} %
\frac{\d}{\d{t}}{z}_l\at{t}&=
J_{l-1}\bat{z\at{t}}
-J_l\bat{z\at{t}}&\text{for}\quad{l}\geq1,
\\%
\label{MBD2}\tag{MBD2} %
J_0\at{z}&=-\sum\limits_{l=1}^{\infty}J_l\at{z},&
\\%
\label{MBD3}\tag{MBD3} %
J_l\at{z}&={\GammaCondL}\,
\at{z_1\,z_l-\ND\at{z}%
\frac{q_l}{q_{l+1}}{z_{l+1}}}&\text{for}\quad{l}\geq1.%
\end{align} %
In what follows we will refer to this system as the \emph{modified
Becker-D\"{o}ring equations}. Moreover, we always assume
\begin{align}%
\tag{A1}\label{ASS1} %
0<R:=\lim\limits_{l\rightarrow\infty}
\frac{q_{l}}{q_{l+1}}<\infty,
\end{align}
as well as
\begin{align}
\label{ASS2}\tag{A2}%
\lim\limits_{l\rightarrow\infty}\frac{\GammaCondL}{l}=0.
\end{align}
Note that \eqref{ASS1} implies the identity
$1/R=\lim_{l\rightarrow\infty}{q_l^{1/l}}$.
\subsubsection*{Aims and results}%
This paper is organized as follows. In Section \ref{secExist} we
give a brief survey on existence and uniqueness results for the
modified equations. We will skip some technical details, because in
this part we mainly adapt methods which are well established for the
standard model.
\par%
In Section \ref{secEqui} we investigate equilibrium states for the
dynamical equations. Our first result is a necessary and sufficient
condition \eqref{EQ} for the existence of such equilibrium states.
Since this condition depends only on some properties of the sequence
$\at{a_l}_{l\in\Nset}$, there is no upper bound for the mass of an
equilibrium state. In other words, \eqref{EQ} implies that for all
$\equi\varrho>0$ there exists a unique and nonnegative equilibrium
state $\equi{z}$ with $\varrho\at{\equi{z}}=\equi\varrho$. Moreover,
in Section \ref{secEqui} we study the minimization problem
$A\at{z}\rra\min$ under the constraint $\varrho\at{z}=\equi\varrho$,
where $\equi\varrho>0$ is fixed, and we prove the following two
statements. \textit{1}. If \eqref{EQ} is satisfied, then the
equilibrium state with mass $\equi\varrho$ is a minimizer.
\textit{2}. In the case that \eqref{EQ} is violated there is no
minimizer at all, but the infimum is $\equi\varrho\,\ln{R}$.
\par%
Section \ref{SecLimit} is devoted to the limit $t\rra\infty$, where
the main problem is the following. Although the mass is conserved
for finite times, see Section \ref{secExist}, some amount of mass
may disappear in the limit $t\rra\infty$. At first we show that for
$t\rra\infty$ the state $z\at{t}$ converges (in some weak sense)
either to an equilibriums state with positive mass or to $0$.
Second, we state and prove an sufficient condition for that the
mass remains conserved for $t\rra\infty$. Finally, we identify
several cases, and prove for most of them that either all mass
is conserved or all mass disappears.

%% file: ch_exist.tex
\section{Existence and Uniqueness}\label{secExist}%
Our main goal within this section is to prove the global existence
of nonnegative, \emph{weak} solutions for the initial value problem
of \eqref{MBD1}--\eqref{MBD3}, see Theorem \ref{ExistTheo3} below.
Furthermore, we will explain how uniqueness results can be derived.
For these reasons we fix some nonnegative initial data $\tilde{z}$
with ${\varrho}_0:=\varrho\at{\tilde{z}}>0$ and $\tilde{z}_l\geq{0}$
for all $l\in\Nset$, and for simplicity we assume $\tilde{z}_1>0$.
We seek solutions $t\mapsto{}z\at{t}$ of the Becker-D\"{o}ring
equations in the space $C\at{\cointerval{0}{\infty};\,X}$, where
the \emph{state space} $X$ is given by
\begin{align}
\label{StateSpace}
X&=\Big\{%
z=\at{z_l}_{l\in\Nset}\;:\;\norm{z}_X<\infty%
\Big\},\quad \norm{z}_X=\sum\limits_{l=1}^{\infty}\,l\,\abs{z_l},
\end{align}
Since we are only interested in solutions of the Becker-D\"{o}ring
equations which are positive or at least nonnegative, we introduce
the cones $X_{0+}$ and $X_{+}$ of all nonnegative and strictly
positive, respectively, elements of $X$, {\ie}
\begin{align}%
X_{0+}=
\Big\{%
z\in{X}\,:\,z_l\geq{0}\;\;\forall\;{l}\in\Nset%
\Big\},\quad%
X_{+}=
\Big\{%
z\in{X}\,:\,z_l>{0}\;\;\forall\;{l}\in\Nset%
\Big\}.
\end{align}
We cite some results of \cite{BCP86}.
\begin{prop}[Ball, Carr, Penrose]%
\label{XSpaceProperties}%
The space $X$ is a Banach space, and it is the dual space of
\begin{align}
\label{PreDualStateSpace}
{}^{\star\!}{X}&=\Big\{%
z=\at{z_l}_{l\in\Nset}\;:\;l^{-1}z_l\xrightarrow{\;l\rra\infty\;}{0}%
\Big\}.
\end{align}
Moreover, let $Z=\at{m\mapsto{\discr{z}{m}}}$ be any sequence in
$X$, and let $\discr{z}{\infty}$ be some element of $X$. Then
\begin{enumerate}
\item%
$Z$ converges to $\discr{z}{\infty}$ weak$^\star$ in $X$ if and
only if
\begin{enumerate}
\item%
the sequence $m\mapsto\norm{\discr{z}{m}}_X$ is bounded, and
\item%
\begin{math}%
\D\discr{z_l}{m}%
\xrightarrow{\;m\rightarrow\infty\;}\discr{z_l}{\infty}
\end{math}
for all
$l\in\Nset$. %
\end{enumerate}
\item%
$Z$ converges to $\discr{z}{\infty}$ strongly in $X$ if and only
if
\begin{enumerate}
\item
\begin{math}%
\discr{z}{m}%
\xrightarrow{\;m\rightarrow\infty\;}%
\discr{z}{\infty}
\end{math}
weak$^\star$ in $X$, and
\item %
$\norm{\discr{z}{m}}_X%
\xrightarrow{\;m\rightarrow\infty\;}%
\norm{\discr{z}{\infty}}_X$.
\end{enumerate}
\end{enumerate}
\end{prop}
\emph{Remarks}. $\at{i}$ The flux $J_l$ is always weak$^\star$ continuous for
$l\geq{1}$. $\at{ii}$ Assumption \eqref{ASS2} provides the weak$^\star$
continuity of $J_0$. $\at{iii}$ Assumption \eqref{ASS1} implies that the
sequence $l\mapsto\abs{\,l^{-1}\ln{q_l}}$ is bounded, and the
availability functional $A$ from \eqref{FreeEnergyNewModel3} is thus
well defined on the whole cone $X_{0+}$.
$\at{iv}$ The cone $X_{0+}$ is closed under both strong and weak$^\star$
convergence, and with \eqref{IntroMassDef} we have
$\norm{z}_X=\varrho\at{z}$ for all $z\in{}X_{0+}$.
\par%
For later purposes we define weak$^\star$ continuous functionals $N_l$, $l\geq{1}$, by
\begin{align}
N_l\at{z}:=\sum\limits_{n=l}^{\infty}z_n.
\end{align}
Clearly, this definition implies $\ND\at{z}=N_1\at{z}$ and
$z_l=N_l\at{z}-N_{l+1}\at{z}$. Moreover, by means of formal
transformations we find
$\varrho\at{z}=\sum_{l=1}^{\infty}N_l\at{z}$ and
\begin{align}
\frac{\d}{\d{t}}\NL_l\bat{z\at{t}}=%
J_{l-1}\bat{z\at{t}}\quad\text{for all}\;{l}\in\Nset\cup\{0\}.
\end{align}
The existence of solutions for the modified model can be proved
similarly to the classical results in \cite{BCP86}:  In the first
step we consider a finite, $m$-dimensional approximate problem,
which results from the infinite system by neglecting all droplets
with more than $m$ atoms. This gives rise to the following system of ordinary
differential equations
\begin{align}
\begin{split}
\frac{\d}{\d{t}}{\discr{{z}_l}{m}}\at{t}&=
J_{l-1}\at{\discr{z}{m}\at{t}}
-J_l\at{\discr{z}{m}\at{t}},\quad{l}=2,\,\tdots,\,m-1,\\
\frac{\d}{\d{t}}{\discr{{z}_m}{m}\at{t}}&=
J_{m-1}\at{\discr{z}{m}\at{t}},\\
\frac{\d}{\d{t}}{\discr{{z}_1}{m}\at{t}}&=
-J_{1}\at{\discr{z}{m}\at{t}}-%
\sum\limits_{l=1}^{m-1}J_l\at{\discr{z}{m}\at{t}},
\end{split}
\label{BDSystemFinSolution} %
\end{align} %
with initial condition
\begin{align}
\label{BDSystemFinSolutionIni} %
{\discr{{z}_l}{m}}\at{0}&=%
\discr{\tilde{z}_l}{m},\quad%
\quad\discr{\tilde{z}_l}{m}= \tilde{z}_l\quad\text{for}\;l=1,\,{\tdots},\,m.
\end{align} %
In the second step we construct weak solutions of the infinite
system \eqref{MBD1}--\eqref{MBD3} as weak$^\star$ limits of solutions to
\eqref{BDSystemFinSolution}--\eqref{BDSystemFinSolutionIni}. 
\par
\emph{Remarks}. $\at{i}$ The vector $\discr{z}{m}$ can be regarded
as an element of $X$ by setting $\discr{{z}_l}{m}\equiv0$ for all
$l>m$. $\at{ii}$ The approximate system is again closed by
\eqref{MBD3}. $\at{iii}$ The initial data $\discr{\tilde{z}}{m}$ of
the approximate system converge for $m\rightarrow\infty$ strongly in
$X$ to $\tilde{z}$, the initial data of the infinite system.
\par%
Existence and uniqueness results for the finite dimensional
{I}{V}{P}
\eqref{BDSystemFinSolution}--\eqref{BDSystemFinSolutionIni} can be
established by means of standard techniques for {ODE}s.
\begin{lem}%
\label{ExistLemma1}%
For all $m\in\Nset$ there exists a smooth and nonnegative solution
$\discr{z}{m}\in{C^\infty}\at{\cointerval{0}{\infty};\,X}$ of the
approximate IVP
\eqref{BDSystemFinSolution}--\eqref{BDSystemFinSolutionIni}.
Moreover, with
\begin{align}
\discr{\ND_l}{m}=%
\ND\at{\discr{z}{m}},\;\discr{{J}_l}{m}=J_l\at{\discr{z}{m}},\;
\discr{A}{m}=A\at{\discr{z}{m}},\;%
\discr{\varrho}{m}=\varrho\at{\discr{z}{m}}
\end{align}
we find
$\discr{{\varrho}}{m}\at{t}=\discr{{\varrho}}{m}\at{0}$ and
\begin{align}
\label{FinFreeEnergyProd}
-\frac{\d}{\d{t}}\discr{{A}}{m}\at{t}\geq%
\frac{\const}{\D\at{\discr{\varrho}{m}}^2}%
\sum\limits_{l=1}^{m-1}%
\abs{\discr{J_l}{m}\at{t}}^2%
\quad\text{with $\const>0$},
\end{align}
for all $t\geq{0}$, and
$\D\frac{\d}{\d{t}}\discr{{\NL}_l}{m}\at{t}=\discr{J_{l-1}}{m}\at{t}$ for all
$t\geq{0}$ and all ${l}=1,\,{\tdots},\,m$.
\end{lem}
\begin{pf}%
For brevity we prove only \eqref{FinFreeEnergyProd}. With similar
transformations as in \eqref{FreeEnergyNewModel4} and exploiting
\eqref{MBD3} we obtain
\begin{align}%
\notag%
\frac{\d}{\d{t}}{A}\at{z}&=
\sum\limits_{l=1}^{m-1}\,\GammaCondL\,%
\at{\discr{z_1}{m}\,\discr{z_l}{m}-%
\discr{\NL_1}{m}\,\frac{q_l}{q_{l+1}}\,%
\discr{z_{l+1}}{m}}\,%
\ln\at{\frac{q_l\,\discr{z_{l+1}}{m}\,%
\discr{\NL_1}{m}}{q_{l+1}\,%
\discr{z_l}{m}\,\discr{z_1}{m}}}
\\&=
-\sum\limits_{l=1}^{m-1}\,\GammaCondL\,%
\bat{d_l-c_l}\,\bat{\ln{c_l}-\ln{d_l}}%
\end{align}
with $c_l=\discr{z_1}{m}\,\discr{z_l}{m}$ and
$d_l=\discr{\NL_1}{m}\,\discr{z_{l+1}}{m}\,q_l/q_{l+1}$. From $\lim_{l\rra\infty}{q_l}/q_{l+1}=R$ and
\begin{align}
\discr{z_l}{m}\leq\discr{\varrho}{m}/l%
\quad\text{and}\quad
\discr{z_1}{m}\leq{\discr{\NL_1}{m}}\leq\discr{\varrho}{m}
\end{align}
it follows that $c_l$, $d_l<{\const}\,\at{\discr{\varrho}{m}}^2/l$,
and hence
\begin{align}
\notag%
\GammaCondL\bat{c_l-d_l}\,
\bat{\ln{c_l}-\ln{d_l}}&%
\geq\frac{\GammaCondL}%
{\max\{c_l,\,d_l\}}\at{c_l-d_l}^2
\\\notag
&\geq
\frac{l\,\GammaCondL}%
{\const\,\D\at{\discr{\varrho}{m}}^2}\at{c_l-d_l}^2
\\\label{FinFreeEnergyProdAA}%
&\geq
\frac{\const}{\D\at{\discr{\varrho}{m}}^2}%
\frac{l}{\GammaCondL}\abs{\discr{J_l}{m}}^2.
\end{align}
Assumption  \eqref{ASS2} implies $l/\GammaCondL\geq\const>0$, and
\eqref{FinFreeEnergyProd} follows from \eqref{FinFreeEnergyProdAA}.
\qed\end{pf}%
In order to pass to the limit $m\rightarrow\infty$ we need some
uniform estimates for the solution of the approximate problem.
\begin{lem}%
\label{ExistLemma2} The following functions in Lemma \ref{ExistLemma1}
are uniformly, {\ie} independently of $m$, bounded in
$C\at{\cointerval{0}{\infty}}$.
\begin{enumerate}
\item %
        $\discr{z_l}{m}$,
        $\discr{\NL_l}{m}$,
        $\discr{J_l}{m}$,
        $\discr{\dot{z}_l}{m}$, and
        $\discr{\dot{\NL}_l}{m}$ for all
        $l\geq{1}$,
\item %
        $\discr{\ddot{z}_l}{m}$,
        $\discr{\ddot{\NL}_l}{m}$,
        $\discr{\dot{J}_l}{m}$ for all
        $l\geq{2}$,
\item %
        $\discr{J_0}{m}$,
        $\discr{\dot{J}_1}{m}$.
\end{enumerate}
\end{lem}
For brevity we omit the proof, which is carried out  in
\cite{Nal05}. Moreover, we can derive all assertions quite easily
from the equations \eqref{BDSystemFinSolution} and assumption
\eqref{ASS2}.
\begin{thm} %
\label{ExistTheo3}%
Let $\discr{z}{m}$ as in Lemma  \ref{ExistLemma1}. Then there
exists a subsequence $j\mapsto\discr{z}{m_j}$, and a function
${z}\in{C}\at{I;\,X}$, $I=\cointerval{0}{\infty}$, with the
following properties.
\begin{enumerate}
\item
    \begin{enumerate}
    \item The convergences
          \begin{math}
          \discr{z_l}{m}
          \xrightarrow{\;j\rightarrow\infty\;}
          z_l
          \end{math}
          and
        \begin{math}
          \discr{\ND_l}{m}
          \xrightarrow{\;j\rightarrow\infty\;}
          \ND_l\at{z}
          \end{math}
          are
          strong in ${C}\at{I}$ for $l=1$, and
          and even strong in $C^1\at{I}$ for $l\geq{2}$,
    \item The convergence
        \begin{math}
        \discr{J_l}{m}\xrightarrow{\;j\rightarrow\infty\;}
        J_l\at{z}
        \end{math}
        is strong in ${C}\at{I}$ for $l=0$, and
        even strong in ${C^1}\at{I}$ for $l\geq{1}$.
    \end{enumerate}
\item   We have $z_l\at{t}\geq{0}$
        for all $l\geq{1}$ and all $t\in{I}$,
\item
        The limit $z$ satisfies
        \begin{align}
            \label{ExistEqn3}
            \frac{\d}{\d{t}}z_l\at{t}=J_{l-1}\bat{z\at{t}}-%
	    J_l\bat{z\at{t}},\quad
            \frac{\d}{\d{t}}N_l%
            \bat{z\at{t}}=J_{l-1}\bat{z\at{t}},
        \end{align}
    for all $l\geq{2}$ and all
    $t\in\cointerval{0}{\infty}$, and
    for all $t_1$, $t_2\in{I}$ we have
        \begin{align}
            \label{ExistEqn1}
            z_1\at{t_2}-z_1\at{t_1}=
            \int\limits_{t_1}^{t_2}J_0\bat{z\at{t}}\,\d{t}.
        \end{align}
\item%
The availability $A$ decreases according to
\begin{align}
\label{FreeEnergyProd}
{A}\bat{z\at{t_1}}- {A}\bat{z\at{t_2}}%
\geq\frac{\const}{\at{\varrho_0}^2}%
\int\limits_{t_1}^{t_2}\sum\limits_{l=1}^{\infty}
\abs{J_l\bat{z\at{t}}}^2\d{t}\geq{0}.
\end{align}
\end{enumerate}%
\end{thm}
\begin{thm}
\label{ExistTheo4}%
The total mass of $z$ from Theorem \ref{ExistTheo3} is conserved,
{\ie} $\varrho\at{z\at{t}}=\varrho\at{z\at{0}}=\varrho_0$ for all
finite $t\geq{0}$.
\end{thm}

\textbf{Remarks}.
\begin{enumerate}
\item %
Because of \eqref{ExistEqn1} the limit $z$ is a \emph{weak} solution
of \eqref{MBD1}--\eqref{MBD3}.
\item %
In Section \ref{SecLimit} it turns out to be useful that
\eqref{ExistEqn3} holds in a strong sense for large $l$.
\item %
Inequality \eqref{FreeEnergyProd} follows from
\eqref{FinFreeEnergyProd} and the Lemma of Fatou. All other
assertions of Theorem \ref{ExistTheo3} are consequences of the
uniform bounds in Lemma \ref{ExistLemma2} and the
{{A}{r}{z}{e}l\'{a}}-Ascoli Theorem, see \cite{Nal05}. Moreover, we
obtain uniform continuity with respect to time for several functions
including $z_l$, $\NL_l\at{z}$, $J_l\at{z}$ for $l\geq1$.
\item%
The proof of Theorem \ref{ExistTheo4} is not so obvious and needs
some careful estimates for the mass contained in the tail of the
solution. However, since one can use similar
methods as in \cite{BCP86} we skip the proof and refer to
\cite{Nal05}.
\end{enumerate}
Finally we give a brief summary of the uniqueness results in
\cite{Nal05}. To establish uniqueness for the infinite system
\eqref{MBD1}--\eqref{MBD3} it is convenient to pass to new variables
$\zeta=\at{\zeta_l}_{l\in\Nset}$ with
\begin{align}
\label{ExistZetaDef}
\zeta_l:=\NL_l\at{z}=\sum\limits_{n=l}^{\infty}z_n.
\end{align}
Note that $z_l=\zeta_l-\zeta_{l+1}$, $\ND\at{z}=\zeta_1$, and
$\varrho\at{z}=\sum_{l=1}^\infty{}\zeta_l$. The change of variables
transforms \eqref{MBD1}--\eqref{MBD2} into
\begin{align}
\label{BDSystemEqn4} %
\frac{\d}{\d{t}}{\zeta}_l\at{t}&=
J_{l-1}\bat{\zeta\at{t}}&\text{for}\quad{l}\geq1,
\\%
\label{BDSystemEqn5} %
J_0\at{\zeta}&=-\sum\limits_{l=1}^{\infty}J_l\at{\zeta}&
\\%
\label{BDSystemEqn6} %
J_l\at{\zeta}&={\GammaCondL}\,
\Bat{\at{\zeta_1-\zeta_2}\,\at{\zeta_{l}-\zeta_{l+1}}-\zeta_1%
\frac{q_l}{q_{l+1}}\at{\zeta_{l+1}-\zeta_{l+2}}}&\text{for}\quad{l}\geq1.%
\end{align} %
Note that Theorems \ref{ExistTheo3} and \ref{ExistTheo4} yield the
global existence of weak solutions for
\eqref{BDSystemEqn4}--\eqref{BDSystemEqn6}. The reformulation of the
original system now provides uniqueness results in form of Gronwall
type estimates.
\begin{thm}%
\label{ExistTheo5}
Let $\discr{\zeta}{1}$ and $\discr{\zeta}{2}$ be two
weak solutions of \eqref{BDSystemEqn4}--\eqref{BDSystemEqn6},
and set $\tilde{\zeta}=\discr{\zeta}{2}-\discr{\zeta}{1}$.
Then there exists a time dependent constant $C\at{t}$ such that
\begin{align}
\notag%
\norm{\tilde{\zeta}\at{t}}_{\ellOne}%
&\leq\norm{\tilde{\zeta}\at{0}}_{\ellOne}
+
{C\at{t}}\,\int\limits_{0}^{t}%
\norm{\tilde{\zeta}\at{s}}_{\ellOne}\,\d{s}.
\end{align}
\end{thm}
A similar result for the classical Becker-D\"{o}ring equations is
derived in \cite{LM02}, and the basic estimates therein can easily
be adapted for proving Theorem  \ref{ExistTheo5}.
This is done in \cite{Nal05}.%

%% file: ch_equi.tex
\section{Equilibrium states}%
\label{secEqui} %
An equilibrium state of the Becker-D\"{o}ring system is a state
$\equi{z}\in{X_+}$, such that all fluxes $J_l$ vanish in $\equi{z}$.
Clearly, $0\in{X}$ is always an equilibrium state. In this section
we study equilibrium states with prescribed positive total mass
$\varrho\at{\equi{z}}=\equi\varrho$. Here $\equi\varrho>0$ is a
given constant which remains fixed within this section.
\par%
For the analysis it is convenient to use the following variant
$\tilde{A}$ of the availability
\begin{align}%
\label{TildeFreeEnergy}
\tilde{A}\at{z}&=A\at{z}-\varrho\at{z}\ln{R}=
\sum\limits_{l=1}^{\infty}z_l
\ln\at{\frac{z_l}{\tilde{q}_l\ND\at{z}}},
\end{align}%
with $\tilde{q}_l=q_l{R^l}$ and $R$ as in \eqref{ASS1}, because
$\tilde{A}$ is weak$^\star$ continuous on $X_{0+}$. To prove this,
we split $\tilde{A}$ into three parts
$\tilde{A}=\tilde{A}_1+\tilde{A}_2+\tilde{A}_3$, where
\begin{align}%
\label{TildeFreeEnergyEqn1}
\tilde{A}_1\at{z}&=-\ND\at{z}\ln\at{\ND\at{z}},\\%
\label{TildeFreeEnergyEqn2}
\tilde{A}_2\at{z}&=\sum\limits_{l=1}^{\infty}z_l\ln{z_l},\\%
\label{TildeFreeEnergyEqn3}
\tilde{A}_3\at{z}&=-\sum\limits_{l=1}^{\infty}z_l\ln{\tilde{q}_l}.%
\end{align}%
The weak$^\star$ continuity of $\tilde{A}_1$ is obvious, of
$\tilde{A}_2$ it was proved in \cite{BCP86}, and of $\tilde{A}_3$
it is a consequence of Proposition \ref{XSpaceProperties} and
\begin{math}%
\lim_{l\rightarrow\infty}l^{-1}\ln{\tilde{q}_l}=\ln{1}=0.
\end{math}%
\par%
Next we derive a necessary condition for the existence of an
equilibrium state $\equi{z}$ with prescribed total mass
$\varrho\at{\equi{z}}=\equi{\varrho}>0$. We set $J_l\at{\equi{z}}=0$
in \eqref{MBD3}, and obtain
\begin{align}
\label{EquiCond1a}
\equi{z}_{l+1}=%
\frac{\equi{z}_{1}}{\equi{\ND}}\,%
\frac{q_{l+1}}{q_l}\,\equi{z}_{l}=
\frac{\equi{z}_{1}}{R\,\equi{\ND}}\,%
\frac{\tilde{q}_{l+1}}{\tilde{q}_l}\,\equi{z}_{l},
\end{align}
where $\equi{\ND}=\ND\at{\equi{z}}$. With $\tilde{q}_1=R\,q_1=R$ and
the abbreviation $\equi{\mu}:={\equi{z}_1}/\at{R\,\equi{\ND}}$,
$\equi{\mu}\in\ccinterval{0}{1/R}$, equation \eqref{EquiCond1a}
yields
\begin{align}
\label{EquiCond1}
\equi{z}_l=%
\at{\frac{\equi{z}_1}{R\,\equi{N}}}^{l-1}\,%
\frac{\tilde{q}_l}{\tilde{q}_1}\,\equi{z}_1=%
\equi{\ND}\,{\tilde{q}_l}\,\equi{\mu}^l=
\equi{\ND}\,{{q}_l}\,\at{R\,\equi{\mu}}^l%
\quad\quad\text{for all}\;{l}\in\Nset.
\end{align}
Finally, the condition $\equi{\ND}=\ND\at{\equi{z}}$ and the
constraint $\varrho\at{\equi{z}}=\equi{\varrho}$ require
\begin{align}
\notag
\equi{\ND}\sum\limits_{l=1}^{\infty}%
\tilde{q}_l\,\equi{\mu}^l=\equi{\ND}%
\quad\text{as well as}\quad
\equi{\ND}\sum\limits_{l=1}^{\infty}%
l\,\tilde{q}_l\,\equi{\mu}^l=\equi{\varrho},
\end{align}
which imply
\begin{align}%
\label{EquiCond2}%
\tilde{f}\at{\equi{\mu}}=1%
\quad\text{and}\quad
{\equi{\ND}}= \frac{\equi\varrho}{\tilde{g}\at{\equi\mu}}%
\quad\text{with}\quad
\tilde{f}\at{\mu}=
\sum\limits_{l=1}^{\infty}\tilde{q}_l\,\mu^l,\quad
\tilde{g}\at{\mu}=
\sum\limits_{l=1}^{\infty}\tilde{q}_l\,l\,\mu^l.
\end{align}
Note that both power series in \eqref{EquiCond2} have the same
radius of convergence \mbox{$\tilde{R}=1$}. The function $\tilde{f}$
is continuous and strictly increasing on $\ccinterval{0}{1}$, and
satisfies $\tilde{f}\at{\mu}\geq\tilde{q_1}\,\mu=\mu\,R$.
Consequently, the parameter $\equi{\mu}$ exists in the interval
$\ccinterval{0}{\min\{1,\,1/R\}}$ if and only if
$\tilde{f}\at{1}=\lim_{\mu\rra1}\tilde{f}\at{\mu}\geq{1}$. Moreover,
in order to guarantee $\varrho\at{\equi{z}}=\equi{\varrho}>0$ we
must have $\equi{N}>0$, or equivalently,
$\tilde{g}\at{\equi{\mu}}<\infty$. Since $\tilde{f}\at{1}>1$ implies
$\equi{\mu}<1$ and therefore $\tilde{g}\at{\equi{\mu}}<\infty$, we
end up with the following condition \eqref{EQ} for the existence of
an equilibrium state
\begin{align}%
\label{EQ}\tag{EQ}%
\begin{minipage}{0.8\textwidth}
\begin{math}
\quad\quad
\tilde{f}\at{1}>1,\quad\quad\quad\text{or}\quad\quad\quad%
\tilde{f}\at{1}=1\;\;\text{and}\;\;\tilde{g}\at{1}<\infty.
\end{math}
\end{minipage}
\;\;\,
\end{align}
Its negation reads
\begin{align}%
\label{NEQ}\tag{NEQ}%
\begin{minipage}{0.8\textwidth}
\begin{math}
\quad\quad
\tilde{f}\at{1}<1,\quad\quad\quad\text{or}\quad\quad\quad%
\tilde{f}\at{1}=1\;\;\text{and}\;\;\tilde{g}\at{1}=\infty.
\end{math}
\end{minipage}
\end{align}
\begin{thm}
\label{OptProbTheo3}%
For any $\equi{\varrho}>0$ there exists an equilibrium state
$\equi{z}$ with $\varrho\at{\equi{z}}=\equi{\varrho}$ if and only if
\eqref{EQ} is satisfied.
Moreover, if \eqref{EQ} is satisfied then
\begin{enumerate}
\item[(a)]%
there exists a unique value
$\equi{\mu}\in\ocinterval{0}{1}$ such that
\begin{align}%
\label{OptProbEqn9}%
\tilde{f}\at{\equi{\mu}}=1,
\end{align}
\item[(b)]
$\equi{z}\in{X}_+$ is given as in
\eqref{EquiCond1}--\eqref{EquiCond2}, {\ie}
\begin{align} %
\label{OptProbEqn10}%
\equi{z}_l=\equi{\ND}\,\tilde{q}_l\,\equi{\mu}^{\,l},\quad
\equi{\ND}=\ND\at{\equi{z}}=
\equi{\varrho}/{\tilde{g}\at{\equi{\mu}}},
\end{align}
\item[(c)]
we have
\begin{math} %
\tilde{A}\at{\equi{z}}=\equi{\varrho}\,\ln{\equi{\mu}}\leq0.
\end{math}
\end{enumerate}
\end{thm}
For the two examples from Section \ref{secIntro} the equilibrium 
condition \eqref{EQ} reads as follows. \textit{Example
1}. Equation \eqref{BeDeEx1Eqn1} implies
$R=\exp\at{-\delta}=\tilde{q}_1<1$ and
\begin{align}
\tilde{f}\at{1}&=\exp\at{-\delta}+
\sum\limits_{l=2}^{\infty}\exp\at{-\gamma{l}^{2/3}}%
\notag\\&\leq%
\exp\at{-\delta}+\frac{1}{\gamma^{3/2}}%
\int\limits_{1}^\infty\exp\at{-s^{2/3}}
\mathrm{d}s.%
\end{align}
In particular, for large values\footnote{%
See \cite{DD05} for physically relevant values.
} of both $\delta$ and $\gamma$ there is no equilibrium state
$\equi{z}$. \mbox{\textit{Example 2}}. From \eqref{BeDeEx2Eqn1} we
deduce $R=\exp\at{+\beta}>1$, $\tilde{q_{l}}=1$ for large $l$, and
$\tilde{f}\at{1}=\infty$, so that there always exists the
equilibrium state \eqref{OptProbEqn9}--\eqref{OptProbEqn10} with
$\equi{\mu}<1/R<1$.
\par%

Let $\partial_{z}\tilde{A}\at{z}$ and $\partial_{z}\varrho\at{z}$
denote the Gateaux differentials of $A$ and $\varrho$ in $z$,
respectively, which are well defined for strictly positive
$z\in{X}_+$. By means of basic calculus we derive from
\eqref{OptProbEqn10} that
\begin{align}
\partial_{z}\tilde{A}\at{\equi{z}}=
\at{\ln\frac{\equi{z}_l}{\tilde{q}_l\,\equi{\ND}}}_{l\in\Nset}=
\at{\ln\equi{\mu}^{\,l}}_{l\in\Nset},%
\quad\partial_{z}\varrho\at{\equi{z}}=\at{l}_{l\in\Nset},
\end{align}
and conclude that \eqref{OptProbEqn10} is equivalent to
\begin{align}
\label{EquiCond5}
\partial_{z}\tilde{A}\at{\equi{z}}&=
\at{\ln\equi{\mu}}\,\partial_{z}\varrho\at{\equi{z}}.
\end{align}
However, since the functional $\tilde{A}$ is not convex, it is not
obvious that \eqref{EquiCond5} defines a minimizer of $\tilde{A}$
under the constraint of prescribed mass. For this reason we study
the optimization problem
\begin{align}
\label{OPT}\tag{OPT}%
\tilde{A}_{\min}=\inf\Big\{
\tilde{A}\at{z}\;:\;z\in{X_{0+}},\;\varrho\at{z}=
\equi{\varrho}\Big\}%
\end{align}%
in more detail. Our
main results are formulated in the next two theorems.
\begin{thm} \label{OptProbTheo1}%
For \eqref{EQ} the infimum
$\tilde{A}_{\min}$ in \eqref{OPT} is attained.
Moreover, a minimizer is given by equations
\eqref{OptProbEqn9}--\eqref{OptProbEqn10}.
\end{thm}
\begin{thm} \label{OptProbTheo2}%
For \eqref{NEQ} we have
$\tilde{A}_{\min}=0$ in \eqref{OPT}, but there is no minimizer.
\end{thm}
\subsection{Proof of Theorem \ref{OptProbTheo1}}
\begin{lem}\label{OptProbLemma3}
For $z\in{X}_{0+}$ and any $\mu\in\oointerval{0}{1}$ we have
\begin{align} %
\tilde{A}\at{z}\geq\varrho\at{z}\ln{\mu}-
\ND\at{z}\ln\at{\tilde{f}\at{\mu}}.
\end{align}
\end{lem}%
\begin{pf}%
It is sufficient to consider $z\neq{0}$, so that $\ND\at{z}>0$. At
first we rewrite $T:=\tilde{A}\at{z}-\varrho\at{z}\ln{\mu}$ as follows
\begin{align} \notag%
T&= \tilde{A}\at{z}-\sum\limits_{l=1}^{\infty}z_l\ln\at{\mu^l}=
\sum\limits_{l=1}^{\infty}z_l
\ln\at{\frac{z_l}{\tilde{q}_l\,\ND\at{z}\,\mu^l}}%
\\\notag&=%
\ND\at{z}\,\sum\limits_{l=1}^{\infty}
\at{\tilde{q}_l\,\mu_l}\at{\frac{z_l}
{\tilde{q}_l\,\ND\at{z}\,\mu_l}}
\ln\at{\frac{z_l}{\tilde{q}_l\,\ND\at{z}\,\mu^l}}%
\\&=%
\ND\at{z}\,\at{\D\sum\limits_{l=1}^{\infty}p_l}\,%
\at{\D\sum\limits_{l=1}^{\infty}p_l\,h\at{y_l}}%
/\at{\D\sum\limits_{l=1}^{\infty}p_l},
\end{align}
where $h\at{y}=y\ln{y}$, $p_l=\tilde{q_l}\,\mu^l$, and
$y_l=z_l/\at{\tilde{q}_l\,\ND\at{z}\,\mu^l}$. Note that
\begin{align}%
\sum\limits_{l=1}^{\infty}p_l=\tilde{f}\at{\mu}<\infty,%
\end{align}
and $p_l>0$ for all $l$. Since the function $h$ is convex, Jensen's
inequality yields
\begin{align}
T&\geq\notag%
\ND\at{z}\,\at{\D\sum\limits_{l=1}^{\infty}p_l}\,%
h\at{\at{\D\sum\limits_{l=1}^{\infty}p_l\,{y_l}}/
{\at{\D\sum\limits_{l=1}^{\infty}p_l}}}
\\&=\notag%
\ND\at{z}\,\at{\D\sum\limits_{l=1}^{\infty}p_l}\,%
h\at{\at{\D\sum\limits_{l=1}^{\infty}z_l/{\ND\at{z}}}/
{\at{\D\sum\limits_{l=1}^{\infty}p_l}}}
\\&=%
\ND\at{z}\,\tilde{f}\at{\mu}\,%
h\at{{1}/\tilde{f}\at{\mu}}=-\ND\at{z}\ln\at{\tilde{f}\at{\mu}},
\end{align}
and the proof is complete.
\qed\end{pf}%
\begin{cor}\label{OptProbCorr4}
Suppose \eqref{EQ}, and let $z\in{X}_{0+}$ with
$\varrho\at{z}=\equi\varrho$. Then,
\begin{align}
\tilde{A}\at{z}\geq\equi\varrho\,\ln\at{\equi{\mu}}=
\tilde{A}\at{\equi{z}},
\end{align}
where $\equi{\mu}$ and $\equi{z}$ as in Theorem
\ref{OptProbTheo3}.
In particular, Theorem \ref{OptProbTheo1} is
proved.
\end{cor}%
\begin{pf}%
Set $\mu=\equi{\mu}$ in Lemma \ref{OptProbLemma3}, and compare
with (c) in Theorem \ref{OptProbTheo3}.
\qed\end{pf}%
\subsection{Proof of Theorem \ref{OptProbTheo2}} %
In this section we consider the case $\eqref{NEQ}$, i.e. we assume
either $\tilde{f}\at{1}<1$ or $\tilde{f}\at{1}=1$ and
$\tilde{g}\at{1}=\infty$, and we prove that now the optimization
problem \eqref{OPT} has no minimizer. Recall that
\begin{math}
\lim_{l\rightarrow\infty}{\tilde{q}_l}^{1/l}=1,
\end{math}
and note that $\tilde{f}\at{1}\leq{1}$ implies $\tilde{q}_l\leq1$
for all $l\in\Nset$, as well as
$\lim_{l\rightarrow\infty}{\tilde{q}_l}=0$.
\par%
Our strategy is to construct certain perturbations of $\tilde{q}$,
such that we can rely on the result of the previous section. For
this reason we set
\begin{align}
\Pi &=\Big\{p=\at{p_l}_{l\in\Nset}\;:\;p_l>0\;
\forall\;l\in\Nset,\quad
\limsup\limits_{l\rightarrow\infty}%
\abs{l^{-1}\ln{p_l}}<\infty\Big\},
\end{align}
and define a functional $\mathcal{A}$ an $X_{0+}\times\Pi$ by
\begin{align}
\label{FreeEnergyDef3} \mathcal{A}\pair{z}{p}=
\sum\limits_{l=1}^{\infty}z_l\ln{\frac{z_l}{p_l\,\ND\at{z}}}=
-\ND\at{z}\ln\at{\ND\at{z}}+%
\sum\limits_{l=1}^{\infty}z_l\ln{\frac{z_l}{p_l}},
\end{align}
so that $\tilde{A}\at{z}=\mathcal{A}\pair{z}{\tilde{q}}$ and
$A\at{z}=\mathcal{A}\pair{z}{q}$.  Note that
$\mathcal{A}\pair{z}{p}$ is well defined for all
$\pair{z}{p}\in{X}_{0+}\times\Pi$. Moreover, if
$\lim_{l\rightarrow\infty}{p_l}^{1/l}=1$ the functional
$\mathcal{A}$ is weak$^\star$ continuous with respect to $z$.
\par%
Definition \eqref{FreeEnergyDef3} implies
\begin{align}
\label{FreeEnergyProp1}
\mathcal{A}\pair{z}{\discr{p}{2}}=\mathcal{A}\pair{z}{\discr{p}{1}}+
\sum\limits_{l=1}^{\infty}z_l%
\ln{\frac{\discr{p_l}{1}}{\discr{p_l}{2}}}
\end{align}
where $\discr{p}{1}$, $\discr{p}{2}$ are two arbitrary elements of
$\Pi$. Furthermore, $-\mathcal{A}$ preserves the order in $\Pi$,
{\ie}
\begin{align}
\label{FreeEnergyOrder} \mathcal{A}\pair{z}{\discr{p}{2}}\geq
\mathcal{A}\pair{z}{\discr{p}{1}}\quad\text{for}
\quad{\discr{p}{2}}\leq{\discr{p}{1}}.
\end{align}
\par%
Now we approximate $\tilde{q}$ by a sequence
$\at{m\mapsto\discr{q}{m}}\subset\Pi$, where $\discr{q}{m}$ is defined by
\begin{align}
\label{FreeEnergyApproxDef}
\discr{q}{m}_l&=\max\,\{\tilde{q}_l,\,\pi_m\},\quad
\pi_m=\sup\limits_{l>m}\,\tilde{q}_l.%
\end{align}
Note that $\lim_{m\rightarrow\infty}\pi_m=0$ and that
$0<\pi_m\leq{1}$ for all $m\in\Nset$. If $m$ is large the sequence
$\discr{q}{m}$ is a good approximation of $\tilde{q}$, because both
series differ only for large $l$. In particular,
\begin{align}
\label{FreeEnergyApproxDef2}
l_m:=\min\{l\;:\;\tilde{q}_l\neq\discr{q_l}{m}\}=
\min\{l:\;\tilde{q}_l<\pi_m\}
\quad\xrightarrow{\,m\rightarrow\infty\,}\quad\infty.
\end{align}
If $\tilde{q}$ is a decreasing sequence, as for instance in the
first example from Section \ref{secIntro}, equation
\eqref{FreeEnergyApproxDef2} reduces to
\begin{align}
\discr{q_l}{m}=\tilde{q}_l\quad\text{for}\quad{l\leq{m}},\quad\quad
\discr{q_l}{m}=\tilde{q}_{m+1}\quad\text{for}\quad{l>m}.
\end{align}%
For any $m\in\Nset$ there exists a unique minimizer of
$\mathcal{A}\pair{\cdot}{\discr{q}{m}}$, because we find
\begin{align}
\sum_{l=1}^{\infty}\discr{q}{m}_l=\infty>1\quad\text{and}\quad
\lim_{l\rightarrow\infty}\at{\discr{q_l}{m}}^{1/l}=%
\lim_{l\rightarrow\infty}\at{\pi_{m}}^{1/l}=1,
\end{align}
and thus there exist variants of Theorems \ref{OptProbTheo3} and
\ref{OptProbTheo1} with $\discr{q}{m}$ instead of $\tilde{q}$. This
provides the existence of
\begin{align}%
\discr{{A}_{\min}}{m}:=
\min\Big\{\mathcal{A}\pair{z}{\discr{q}{m}}\;:\;\varrho\at{z}=
\equi\varrho\Big\},
\end{align} as well as the identity
\begin{align} \label{FreeEnergyApproxMin1}
\discr{{A}_{\min}}{m}=\equi{\varrho}\,\ln{\mu_m}=
\mathcal{A}\pair{\discr{z}{m}}{\discr{q}{m}},
\end{align}
where $\mu_m\in\oointerval{0}{1}$ and $\discr{z}{m}\in{X_+}$ satisfy 
\begin{align}
\label{FreeEnergyApproxMin2}
\sum\limits_{l=1}^{\infty}\discr{q_l}{m}\,\mu_m^l=1,
\quad{}\discr{z_l}{m}=N_m \,\discr{q_l}{m}\,{\mu_m}^l,
\end{align}
where
\begin{math}
N_m=\equi{\varrho}
/\at{\sum_{l=1}^{\infty}\discr{q_l}{m}\,l\,\mu_m^l}
\end{math}.
Recall that $\varrho\at{\discr{z}{m}}=\equi{\varrho}$ for all $m$.
\par%
Definition \eqref{FreeEnergyApproxDef} implies
\begin{align}
\label{FreeEnergyChain1} \tilde{q}\leq\tdots%
\leq{}q^{m+1}\leq{q}^m\leq\tdots\leq{q}^1\leq{1}.
\end{align}
This chain and \eqref{FreeEnergyOrder} give
\begin{align}
\label{FreeEnergyChain2} \mathcal{A}\pair{z}{\tilde{q}}%
\geq\tdots
\geq\mathcal{A}\pair{z}{q^{m+1}}
\geq\mathcal{A}\pair{z}{{q}^m}\geq\tdots
\geq\mathcal{A}\pair{z}{q^{1}}\;\;\; \forall\;z\in{X}_{0+},
\end{align}
and hence
\begin{align}
\label{FreeEnergyChain3} \tilde{A}_{\min}%
\geq\tdots
\geq\discr{{A}_{\min}}{m+1}\geq\discr{{A}_{\min}}{m}%
\geq\tdots
\geq\discr{{A}_{\min}}{1},
\end{align}
where \begin{math} \tilde{A}_{\min}=
\inf\Big\{\mathcal{A}\pair{z}{\tilde{q}}\;:\;
\varrho\at{z}=\equi\varrho\Big\}.
\end{math}
\begin{lem}$\;$
\label{FreeEnergyLemma1}
\begin{enumerate}
\item[(a)]
For any $z\in{X}_{0+}$ and $m\rightarrow\infty$ we have
\begin{math}
\mathcal{A}\pair{z}{\discr{q}{m}}
\uparrow\mathcal{A}\pair{z}{\tilde{q}}.
\end{math}%
\item[(b)] %
The sequence $m\mapsto{}\discr{z}{m}$ from
\eqref{FreeEnergyApproxMin2} is a minimizing sequence for
$\tilde{A}$, and
\begin{math}
\discr{{A}_{\min}}{m}\uparrow{}\tilde{A}_{\min}
\end{math}
 for $m\rightarrow\infty$,
\item[(c)]
 $\tilde{A}_{\min}=0$.
\end{enumerate}
\end{lem}
\begin{pf}%
Using \eqref{FreeEnergyProp1} and  $\discr{q_l}{m}\leq1$ we find
\begin{align}
\notag%
\mathcal{A}\pair{z}{\tilde{q}}-
\mathcal{A}\pair{z}{\discr{q}{m}}&=\sum\limits_{l=1}^{\infty}
z_l\ln{\frac{\discr{q_l}{m}}{\tilde{q}_l}} \notag\\&=%
\sum\limits_{l\,:\,\tilde{q}_l\neq{}\discr{q_l}{m}}^{}z_l
\ln{\frac{\discr{q_l}{m}}{\tilde{q}_l}}
\leq%
\sum\limits_{l\,:\,\tilde{q}_l\neq{}\discr{q_l}{m}}^{}z_l
\ln{\frac{1}{\tilde{q}_l}},
\end{align}
and H\"{o}lder's inequality gives
\begin{align}
\notag%
\mathcal{A}\pair{z}{\tilde{q}}-
\mathcal{A}\pair{z}{\discr{q}{m}}&\leq
\at{\sup\limits_{l\,:\,\tilde{q}_l\neq{}\discr{q_l}{m}}
\abs{l^{-1}\ln{\tilde{q}_l}}}
\at{\sum\limits_{l\,:\,\tilde{q}_l\neq{}\discr{q_l}{m}}z_l\,l}
\\&\leq%
\label{FreeEnergyLemma1Eqn2} %
\at{\sup\limits_{l\,:\,\tilde{q}_l\neq{}\discr{q_l}{m}}
\abs{\ln{\tilde{q}_l^{1/l}}}} \at{\sum\limits_{l=1}^{\infty}z_l\,l}
\leq%
\equi{\varrho}\,%
{\sup\limits_{l\geq{l_m}}\abs{\ln{\tilde{q}_l^{1/l}}}},
\end{align}
 where $l_m$ is defined in \eqref{FreeEnergyApproxDef2}.
Combining \eqref{FreeEnergyLemma1Eqn2} and
\eqref{FreeEnergyChain2} yields
\begin{align}
\label{FreeEnergyLemma1Eqn3} %
\mathcal{A}\pair{z}{\discr{q}{m}}&
\leq\mathcal{A}\pair{z}{\tilde{q}}\leq
\mathcal{A}\pair{z}{\discr{q}{m}}+\equi\varrho\,\eta_m,
\end{align}
were $\eta_m$ abbreviates
$\eta_m=\sup_{l\geq{l_m}}\abs{l^{-1}\ln{\tilde{q}_l}}$. The
limit $\lim_{l\rightarrow\infty}\tilde{q}_l^{1/l}=1$ implies
\begin{math}
\lim_{l\rightarrow\infty}\abs{l^{-1}\ln{\tilde{q}_l}}=0,
\end{math}
and thanks to \eqref{FreeEnergyApproxDef2} we find
$\lim_{m\rightarrow\infty}\eta_m=0$. Since
\eqref{FreeEnergyChain2} provides the monotonicity as well as the
convergence of the sequence
$m\mapsto\mathcal{A}\pair{z}{\discr{q}{m}}$ we can pass to the
limit $m\rightarrow\infty$  in \eqref{FreeEnergyLemma1Eqn3}, and
obtain assertion (a). Evaluating \eqref{FreeEnergyLemma1Eqn3} for
$z=\discr{z}{m}$ gives
\begin{align}
\discr{{A}_{\min}}{m}\leq%
\mathcal{A}\pair{\discr{z}{m}}{\tilde{q}}\leq
\discr{{A}_{\min}}{m}+{\equi{\varrho}}\,\eta_m.
\end{align}
Moreover, \eqref{FreeEnergyChain3} implies
\begin{align}
\label{FreeEnergyLemma1Eqn4} %
\discr{{A}_{\min}}{m}\leq{\tilde{A}_{\min}}\leq
\mathcal{A}\pair{\discr{z}{m}}{\tilde{q}}\leq
\discr{{A}_{\min}}{m}+{\equi{\varrho}}\,\eta_m.
\end{align}
Assertion (b) now follows from passing to the limit
$m\rightarrow\infty$ in \eqref{FreeEnergyLemma1Eqn4}. Finally we
prove assertion (c). From \eqref{FreeEnergyChain1} and
\eqref{FreeEnergyApproxMin2}$_1$ we derive
\begin{align}%
\mu_1\leq\tdots\leq\mu_m\leq\mu_{m+1}%
\leq\tdots\leq1 .
\end{align}
Thus there exists
$\tilde{\mu}:=\lim_{m\rightarrow\infty}\mu_m\leq1$. Suppose for
contradiction that $\tilde{\mu}<1$.  Then, %
\begin{align}
\at{m\mapsto\at{\mu_m^l}_{l\in\Nset}}\quad\xrightarrow{\;m\rightarrow\infty\;}\quad
\at{\tilde{\mu}^l}_{l\in\Nset}\quad\quad\text{in}\quad\ellOne.
\end{align}%
Since $\discr{q}{m}$ converges for $m\rightarrow\infty$ to
$\tilde{q}$ in $\ellInf$, we can conclude
\begin{align}
1=\sum\limits_{l=1}^{\infty}\discr{q_l}{m}\,\mu_m^l\quad
\xrightarrow{\,m\rightarrow\infty\,}\quad
\sum\limits_{l=1}^{\infty}\tilde{q}_l\,\tilde{\mu}^l<
\sum\limits_{l=1}^{\infty}\tilde{q}_l\leq1.
\end{align}
This contradiction shows $\tilde{\mu}=1$. Therefore
\begin{align}
{\tilde{A}_{\min}}=\lim\limits_{m\rightarrow\infty}%
\discr{{A}_{\min}}{m}=
\equi{\varrho}\lim\limits_{m\rightarrow\infty}\ln{\mu_m}=0,
\end{align}
which was claimed.
\qed\end{pf}%
\begin{cor}
\label{FreeEnergyCorr3} Since \eqref{EQ} is violated, there is
no minimizer in \eqref{OPT}. In particular, Theorem
\ref{OptProbTheo2} is proved.
\end{cor}
\begin{pf}%
By contradiction. Suppose there is a state $z\in{X}_{0+}$ with
$\varrho\at{z}=\equi{\varrho}>0$ and $\tilde{A}\at{z}=0$. Then
$z\neq0$ and hence $\ND\at{z}>0$. According to Lemma
\ref{OptProbLemma3} we can estimate
\begin{align}
\label{FreeEnergyCorr3Eqn1}%
0={\tilde{A}\at{z}}\geq
{\varrho\at{z}}\ln{\mu}-{\ND\at{z}}\ln{\tilde{f}\at{\mu}} \qquad%
\text{for all $\mu\in\oointerval{0}{1}$}.
\end{align}%
At first suppose $\tilde{f}\at{1}<1$, and let $\mu\rightarrow{1}$.
Then \eqref{FreeEnergyCorr3Eqn1} yields a contradiction, namely
\begin{math}
0={\tilde{A}\at{z}}\geq-{\ND\at{z}}\ln{\tilde{f}\at{1}}>0.
\end{math}
Now suppose $\tilde{f}\at{1}=1$ and $\tilde{g}\at{1}=\infty$. Then,
\eqref{FreeEnergyCorr3Eqn1} implies
\begin{align}
{\varrho\at{z}}\ln{\mu}\leq{\ND\at{z}}\ln{\tilde{f}\at{\mu}}%
\leq{\varrho\at{z}}\ln{\tilde{f}\at{\mu}}
\end{align}
and hence $\mu\leq\tilde{f}\at{\mu}$ for all
$\mu\in\oointerval{0}{1}$. Moreover, from
$\mu\,\tilde{f}^{\prime}\at{\mu}=\tilde{g}\at{\mu}$ we conclude
$\lim_{\mu\rra1}\tilde{f}^{\prime}\at{\mu}=\infty$. Therefore, for
$\mu_1$ and $\mu_2$ with  $\mu_1,\,\mu_2\lesssim1$ we find
\begin{align}
\tilde{f}\at{\mu_2}-\tilde{f}\at{\mu_1}=%
\int\limits_{\mu_1}^{\mu_2}\,%
\tilde{f}^{\prime}\at{\mu}\,\mathrm{d}\mu\geq2\,\at{\mu_2-\mu_1}.
\end{align}
With $\mu_2\rra1$ it follows
\begin{align}
\tilde{f}\at{\mu_1}\leq\tilde{f}\at{1}-2\,\at{1-\mu_1}=2\,\mu_1-1<\mu_1,
\end{align}
which is the desired contradiction.
\qed\end{pf}%

\begin{cor}
Let $m\mapsto{\discr{z}{m}}$ be an arbitrary sequence of
minimizers for problem \eqref{OPT}. Then,
\begin{math}
{\discr{z}{m}}\xrightarrow{\;m\rra\infty\;}0
\end{math} weak$^\star$ in $X_{0+}$.
\end{cor}%
\begin{pf}%
The sequence is bounded and thus weak$^\star$ compact. Let
$j\mapsto{}\discr{z}{m_j}$ be a subsequence, such that
$\discr{z}{m_j}\rightarrow{\discr{z}{\infty}}$ weak$^\star$ in
$X_{0+}$ for $j\rightarrow\infty$. The weak$^\star$  continuity of
$\tilde{A}$ implies $\tilde{A}\at{\discr{z}{\infty}}=0$. Suppose
that $\discr{z}{\infty}\neq0$, {\ie} $\varrho_\infty:=\varrho\at{\discr{z}{\infty}}>0$, and let
$\tilde{z}=\equi{\varrho}\,\discr{z}{\infty}/\varrho_\infty$.
Since
\begin{math}
\tilde{A}\at{\tilde{z}}=
\equi{\varrho}\,\tilde{A}\at{\discr{z}{\infty}}/\varrho_\infty=0,
\end{math}
Corollary \ref{FreeEnergyCorr3} yields a contradiction. We
conclude $\discr{z}{\infty}=0$, which shows that $0$ is the unique
accumulation point of the sequence. This implies the claimed
convergence.
\qed\end{pf}%

%% file: ch_lim.tex
\section{The limit $t\rightarrow\infty$.}%
\label{SecLimit} In this section we study the longtime behavior of
the solution $t\mapsto{z\at{t}}$ from Section \ref{secExist}. At
first we show that any final limit is an equilibrium state, and then
we investigate whether this state is unique, and whether the mass
remains conserved in the limit $t\rra\infty$. Recall that
$\varrho\at{z\at{t}}=\varrho\at{z\at{0}}=\varrho_0$  holds for all
finite times $t\geq{0}$.
\subsection{Auxiliary result}%
\begin{lem}%
\label{LimLemma1}%
For all $l\geq0$ and $t\rightarrow\infty$ we have
\begin{math}
J_l\at{z\at{t}}\rightarrow0
\end{math}.
\end{lem}
\begin{pf}
Suppose for contradiction that there exist some $\eps>0$, an index
$l_0\geq{1}$, and a sequence $m\mapsto{t_m}$ with
$t_m\rightarrow\infty$ for $m\rightarrow\infty$, such that
\begin{align}
\label{LimLemma1Eqn1}%
\abs{J_{l_0}\bat{z\at{t_m}}}\geq{}2\,\eps\quad\text{for
all}\quad{m}\in\Nset.
\end{align}
The uniform continuity of $t\mapsto{}J_{l_0}\Bat{z\at{t}}$, see the
remarks for Theorems \ref{ExistTheo3} and \ref{ExistTheo4}, imply the existence of
$\tau>0$ with
\begin{align}
\abs{J_{l_0}\bat{z\at{t}}}\geq{}\eps\quad\text{for
all}\;{m}\in\Nset\;\text{and}\;{t}\in\oointerval{t_m}{t_m+\tau}.
\end{align}
By extracting a subsequence, still denoted by $m\mapsto{t_m}$, we
can achieve that $t_m+\tau\leq{}t_{m+1}$ for all $m\in\Nset$.
Estimate \eqref{FreeEnergyProd} now implies
\begin{align}
{A}\bat{z\at{t_m}}- {A}\bat{z\at{t_{m+1}}}%
&\geq\frac{\const}{{\varrho_0}^2}\int\limits_{t_m}^{t_m+1}
\abs{J_{l_0}\bat{z\at{t}}}^2\d{t}\geq\tau\eps,
\end{align}
and hence
\begin{math}
{A}\at{z\at{t_{m}}}\rightarrow-\infty
\end{math} for $m\rightarrow\infty$,
which contradicts either Theorem \ref{OptProbTheo1} or Theorem
\ref{OptProbTheo2}. This proves the assertion for all $l\neq{0}$.
Now suppose $l_0=0$ in \eqref{LimLemma1Eqn1}. Without loss of
generality we can assume that there exists
$\discr{z}{\infty}\in{X}_{0+}$ such that
\begin{align}
\discr{z}{m}\xrightarrow{\;m\rra\infty\;}\discr{z}{\infty}\quad%
\text{weak$^\star$ in $X$},
\end{align}%
which implies
$J_l\at{\discr{z}{m}}\xrightarrow{\;m\rra\infty\;}%
J_l\at{\discr{z}{\infty}}$ for all $l\geq{0}$, because all
functionals $J_l$ are weak$^\star$ continuous, see the remarks for 
Proposition \ref{XSpaceProperties}. In the first part of this proof
we have shown that $J_l\at{\discr{z}{\infty}}=0$ for all
$l\geq{1}$. Finally, we find 
\begin{align}%
J_0\at{\discr{z}{\infty}}=%
-\sum\limits_{l=1}^\infty{J_l\at{\discr{z}{\infty}}}=0,
\end{align}
which is a contradiction for \eqref{LimLemma1Eqn1}.
\qed\end{pf}%
Let $X_{\rm{w}}$ be the space $X$ equipped with with the
weak$^\star$ topology, and let  $\omega$ denote the $\omega$-limit
set of $Z=\{z\at{t}\,:\,t\geq{0}\}$ in $X_{\rm{w}}$, {\ie}
\begin{align}
\notag
\omega=\Big\{%
z\in{X}\;:\;%
z=\text{weak}^{\star}\!\!-\!\!\!\lim_{m\rra\infty}z\at{t_m}%
\text{ for some }%
m\mapsto{t_m}\text{ with }%
\lim_{m\rra\infty}t_m=\infty%
\Big\}.
\end{align}

\begin{cor}
\label{LimCorr2} %
$Z$ is relatively compact in $X_{\rm{w}}$, and $\omega$ contains
at least one equilibrium state $\discr{z}{\infty}\in{}X_{0+}$ with
\begin{align}
\label{LimCorr2Eqn1}
\varrho_\infty:=%
\D\varrho\at{\discr{z}{\infty}}\in\ccinterval{0}{\varrho_0}%
\quad\text{and}\quad%
\D\tilde{A}_\infty:=\tilde{A}\bat{\discr{z}{\infty}}=
\lim\limits_{t\rightarrow\infty}\tilde{A}\bat{z\at{t}}.
\end{align}
Moreover, if $\discr{z}{\infty}$ is unique, then %
\begin{align}
z\at{t}\xrightarrow{\;t\rra\infty\;}%
\discr{z}{\infty}\quad\text{weak$^\star$ in $X$},
\end{align}%
and this convergence is strong in $X$ if and only if
\begin{math}
\varrho_\infty=\varrho_0.
\end{math}
\end{cor}
\begin{pf}
Note that $\lim_{t\rightarrow\infty}\tilde{A}\at{t}$ exists, because
the function $t\mapsto\tilde{A}\at{t}$ is decreasing and bounded.
Moreover, since the total mass is conserved, the set $Z$ is
relatively compact in $X_{\rm{w}}$, and therefore $\omega$ contains
some $\discr{z}{\infty}\in{}X$, which clearly satisfies
$\discr{z}{\infty}\in{}X_{0+}$. Lemma \ref{LimLemma1} shows that
$\discr{z}{\infty}$ is in fact an equilibrium state, and
\eqref{LimCorr2Eqn1} comes from the weak$^\star$ continuity
of $\tilde{A}$. The remaining assertions follow from elementary
topological principles and Proposition \ref{XSpaceProperties}.
\qed\end{pf}%
Next we prove a sufficient condition for the mass conservation in
the limit $t\rra\infty$.
\begin{thm}%
\label{MassConservationTheorem1} %
Suppose that there exist $R^{\,\prime}\in\cointerval{0}{R}$ 
and a time $t_0\geq{0}$ such that
\begin{align}
\label{MassConservationTheorem1Eqn1}
\lambda\at{t}:=z_1\at{t}/\ND\at{t}\leq{}R^{\,\prime}
\end{align}
holds for all $t\geq{t_0}$. Then the mass must be conserved for
$t\rra\infty$, {\ie} $\varrho\at{\discr{z}{\infty}}=\varrho_0$ 
for all $\discr{z}{\infty}$ from Corollary \ref{LimCorr2}.
\end{thm}
Note that \eqref{MassConservationTheorem1Eqn1} is equivalent to
$\Gamma^\vap_l\at{t}\geq\kappa\,\Gamma^\cond_l\at{t}$,
$\kappa=R^{\,\prime}/R$, so that the assumption of Theorem
\eqref{MassConservationTheorem1} implies that for large cluster and
large times the fragmentation process dominates coagulation.
\subsection{Main Results}%
Before we prove Theorem \ref{MassConservationTheorem1} we discuss
its consequences. To this end we consider several cases which are
gathered in the following table.
\begin{center}
\begin{tabular}{lllll}
  Case
  &%
  Conditions
  &%
  Limit $t\rra\infty$
  &%
  Convergence\\
  \hline{} \vspace{-20pt}   \\
  {{\emph{NEQ}}}:
  &%
  \eqref{NEQ}
  &%
  $\discr{z}{\infty}=0$
  &%
  weak$^\star$
  \\
  {{\emph{EQ-1}}}:&
  \eqref{EQ}, $\tilde{f}\at{1}=1$
  &%
  OPEN
  &%
  \\
  {{\emph{EQ-2}}}:&
  \eqref{EQ}, $\tilde{f}\at{1}>1$, $R>1$
  &%
  $\discr{z}{\infty}=\equi{z}\at{\varrho_0}$
  &%
  strong
  \\
  {{\emph{EQ-3}}}:&
  \eqref{EQ}, $\tilde{f}\at{1}>1$, $R\leq1$
  &%

  &%
  \\
  \quad{}{\emph{EQ-3a}}:  &
  $\tdots$, $\tilde{A}_\infty=0$, &
  $\discr{z}{\infty}=0$
  &%
  weak$^\star$\\
  \quad{}{{\emph{EQ-3b}}}:  &
  $\tdots$, $\tilde{A}_\infty<0$, &
  $\discr{z}{\infty}=\equi{z}\at{\varrho_0}$&strong
\end{tabular}
\end{center}%
In the case \textit{{N}{E}{Q}} the solution converges weak$^\star$
to $0$, because there is no equilibrium state with positive mass at
all. This case is actually the most interesting one, because here
all mass is contained in larger and larger clusters when time
increases. The same phenomenon occurs within the standard model if
$\varrho_0$ exceeds the critical value $\varrho_{\mathrm{S}}$, and
in this case the long-time dynamics of the large clusters is
governed by the Lifshitz--Slyozov--Wagner ({L}{S}{W}) equation, see
\cite{Pen97} for a formal derivation, and
\cite{CGPV02,LM02,Nie03,Nie04} for rigorous results. We expect to
find an analogue for the {L}{S}{W} equation, now describing the long
time evolution for case \textit{{N}{E}{Q}} of the modified model.
However, this problem is addressed in a forthcoming paper.
\par%
Let us continue with \textit{{{E}{Q}-2}}. As we will see below, the
condition $R>1$ implies the conservation of mass for $t\rra\infty$
without further assumption, {\ie} $\discr{z}{\infty}$ is the 
unique equilibrium state with mass $\varrho_0$, and all claimed results follow
immediately from Corollary \ref{LimCorr2}.
\par%
Next we consider \textit{{E}{Q}-{3}}. Theorem \ref{OptProbTheo3}
provides a family of equilibrium states $\equi{z}$, which are
parameterized by the total mass $\varrho$, or alternatively,
by the availability $\tilde{A}$. Corollary \ref{LimCorr2} states
$\discr{z}{\infty}=\equi{z}(\tilde{A}_\infty)$, i.e.
$\discr{z}{\infty}$ is the unique equilibrium state with
availability $\tilde{A}_\infty$, so that the uniqueness of
$\tilde{A}_\infty$ implies the uniqueness of $\discr{z}{\infty}$.
Now suppose \textit{{{E}{Q}-3b}}. From $\tilde{A}_\infty<0$ it
follows that $\varrho_\infty\neq0$, and we will see that this
already gives $\varrho_\infty=\varrho_0$, {cf.} Corollary
\ref{MassConservationCorollary3} below. However, in subcase
\textit{{E}{Q}-{3a}} we have $\tilde{A}_\infty=0$, {\ie}
$\discr{z}{\infty}=\equi{z}\at{0}=0$, and thus we conclude that
this subcase is very similar to \textit{{N}{E}{Q}}. Note that 
the initial condition $\tilde{A}\at{z\at{0}}<0$ surely implies 
\textit{{E}{Q}-{3b}}. However, for $\tilde{A}\at{z\at{0}}>0$ 
it may depend on the distribution of the initial mass whether the long time
behavior is governed by \textit{{E}{Q}-{3a}} or
\textit{{E}{Q}-{3b}}.
\par%
Finally we discuss \textit{{E}{Q}-{1}}. Again there exists a family
of possible equilibrium states, see Theorem \ref{OptProbTheo3}.
However, all these equilibrium states have the same availability,
because  $\equi{\mu}=1$  implies
$\tilde{A}\at{\equi{z}}=\equi{\varrho}\ln\equi{\mu}=0$, and it
remains open to establish uniqueness.
\par%
Now we have formulated all results concerning the limit
$t\rra\infty$. In the remaining part we prove Theorem
\ref{MassConservationTheorem1} as well as the following result.
\begin{cor}%
\label{MassConservationCorollary3} %
In the cases {\textit{{E}{Q}-{2}} and {\textit{{E}{Q}-3b}}} we have
$\varrho\at{\discr{z}{\infty}}=\varrho_0$.
\end{cor}
\begin{pf}
For Case ${\textit{EQ-2}}$ we set $t_0=0$ and choose
$R^{\,\prime}\in\oointerval{1}{R}$. Then, Theorem
$\ref{MassConservationTheorem1}$ provides the conservation of mass,
because $\lambda\at{t}$ takes values in $\ccinterval{0}{1}$.
Now suppose Case ${\textit{EQ-3b}}$, and recall that
$\tilde{A}\at{\discr{z}{\infty}}<0$ implies
$\discr{z}{\infty}\neq{0}$, and hence $\ND\at{\discr{z}{\infty}}>0$.
Since $z\at{t}\rightarrow\discr{z}{\infty}$ weak$^\star$ in $X$ for
$t\rightarrow\infty$, we find
\begin{align}
\lambda\at{t}=\frac{z_1\at{t}}{\ND\at{z\at{t}}}%
\xrightarrow{\;t\rightarrow\infty\;}
\frac{\discr{z_1}{\infty}}{\ND\at{\discr{z}{\infty}}}=
\tilde{q}_1\,\equi{\mu}=:\equi{\lambda},
\end{align}
and $\tilde{f}\at{1}>1$ gives $\equi\mu<1$ and $\equi{\lambda}<R$.
Consequently, the assumptions of Theorem
\ref{MassConservationTheorem1} are satisfied for
$R^{\,\prime}\in\oointerval{\equi\lambda}{R}$ and $t_0$ sufficiently
large.
\qed\end{pf}
The proof of Theorem \ref{MassConservationTheorem1} consists of
several non-trivial steps, which we present in the following two
subsections. Before we go into details, we shall briefly
describe the main ideas. At first we recall the quantities $\zeta_l$
from Section \ref{secExist}
\begin{align}
\zeta_l=\NL_l\at{z}=\sum\limits_{n=l}^{\infty}z_n,
\end{align}
such that $\zeta_1=\ND\at{z}$ and
$\varrho\at{z}=\sum_{l=1}^{\infty}\zeta_l$.  In what follows we will
identify certain sequences $\sigma=\at{\sigma_l}_{l\in\Nset}$ for
which
\begin{align}
\label{MassConsEqn6} %
{H}_\sigma\at{t}:=
\max\left\{\frac{\varrho_0}{\sigma_{l_0}},\,%
\sup\limits_{l\geq{l_0+1}}
\frac{\zeta_{l}\at{t}}{\sigma_l}\right\}
\end{align}
decreases with time for $t\geq{t_0}$ and for $l_0$ sufficiently
large. Moreover, some of these sequences $\sigma$ are elements of
$\ellOne$. Consequently, for all $\eps>0$ there exists an index
$l_1$, such that
\begin{math}
\sum_{l=l_1}^{\infty}{\zeta_l}\at{t}\leq{}
H_\sigma\at{t_0}\,\sum_{l=l_1}^{\infty}{\sigma_l}\leq{}\eps
\end{math}
holds true for all $t\geq{t_0}$, and this uniform estimate
implies the conservation of mass for $t\rightarrow\infty$.
\par%
The approach described above is inspired by Ball and Carr
\cite{BC88}, which use a similar idea to prove conservation of mass
within the standard model. Another application of the method from
\cite{BC88} is given in \cite{Can05}.
\subsection{More auxiliary results}%
Let $l_0\in\Nset$ be given, and let $\eta=\at{\eta_l}_{l\in\Nset}$
be any strictly positive sequence with
\begin{align}
\label{ToolsEqn1a} 0<\eta_0:=\sup\limits_{l\geq{l_0}}{\eta_l}<1.
\end{align}
Depending on $l_0$ and $\eta$ we define a set $S=S_{\eta,\,l_0}$ by
\begin{align}%
\label{ToolsEqn1} S:=\left\{\sigma=\at{\sigma_l}_{l\in\Nset}\;:\;
\begin{array}{ll}
\at{i}&\sigma_l
\geq\sigma_{l+1}\geq0\;\text{for all}\;l\in\Nset,\\
\at{ii}&\at{\sigma_{l}-\sigma_{l+1}}\geq%
\eta_l\at{\sigma_{l-1}-\sigma_{l}}\;\text{for
all}\;l\geq{l_0}
\end{array}
\right\}.
\end{align}%
Moreover, let $S_+:=S\setminus\{0\}$
and note that \eqref{ToolsEqn1} provides $\sigma_l>0$
for all $\sigma\in{S_+}$ and all $l\in\Nset$.%
\begin{lem}[Ball, Carr]%
\label{ToolsLemma1}%
The set $S$ is closed under $\at{i}$ addition, $\at{ii}$
multiplication with positive constants, $\at{iii}$ pointwise
convergence of sequences, and $\at{iv}$ taking pointwise infima in
arbitrary subsets.
\end{lem}
\begin{pf}%
$\at{i}-\at{iii}$ are obvious. To prove $\at{iv}$, let $I$ be an
arbitrary index set and \mbox{$I\ni{i}\mapsto\discr{\sigma}{i}$}
be any family in $S$. We set
$\sigma_l=\inf_{i\in{I}}\discr{\sigma_l}{i}$ for all $l\in\Nset$. By
construction,
\begin{align}%
\label{ToolsEqn3}%
&\discr{\sigma_{l}}{i} \geq\sigma_l&
\text{for all}\quad&{i\in{I}},\;{l}\in\Nset,\\
\label{ToolsEqn4}%
&\discr{\sigma_{l}}{i}
\geq\discr{\sigma_{l+1}}{i}\geq0%
&\text{for all}\quad&{i\in{I}},\;{l}\in\Nset,\\
\label{ToolsEqn5}%
&\at{\discr{\sigma_{l}}{i}-\discr{\sigma_{l+1}}{i}}%
\geq\eta_l\at{\discr{\sigma_{l-1}}{i}-\discr{\sigma_{l}}{i}}%
&\text{for all}\quad&{i\in{I}},\;{l}\geq{l_0}.
\end{align}
\eqref{ToolsEqn3} and \eqref{ToolsEqn4} imply
$\discr{\sigma_{l}}{i}\geq\discr{\sigma_{l+1}}{i}\geq\sigma_{l+1}\geq{0}$
and hence $\sigma_{l}\geq\sigma_{l+1}\geq{0}$. \eqref{ToolsEqn3} and
\eqref{ToolsEqn5} yield
\begin{align}\notag
\at{1+\eta_l}\discr{\sigma_l}{i}\;\geq\;
\discr{\sigma_{l+1}}{i}+\eta_l\,\discr{\sigma_{l-1}}{i}
\;\geq\;\sigma_{l+1}+\eta_l\,\sigma_{l-1},
\end{align}
and thus
\begin{math}
\at{1+\eta_l}\sigma_l\geq\sigma_{l+1}+\eta_l\,\sigma_{l-1},
\end{math}
as required.
\qed\end{pf}%
We say, a sequence $\sigma\in{S}$ is a \emph{$S$-majorant} of a
sequence $\xi\in\ellInf$, if $\sigma_l\geq\xi_l$ holds for all
$l\geq{l_0}$.
\begin{cor}\label{ToolsCoor2}
There exists an operator
\begin{math}
\,\widehat{\;}\,:\ellInf\rightarrow{S},
\end{math}
mapping $\xi$ to $\widehat{\xi}$, such that for given $\xi$ the
image $\widehat{\xi}$ is the minimal $S$-majorant of $\xi$. This
reads
\begin{align}
\label{ToolsEqn2} \widehat{\xi}&=
\inf\Big\{\sigma\in{S}\;:\;\sigma_l\geq\xi_l\quad \text{for
all}\quad{}l\geq{l_0}\Big\}.
\end{align}
\end{cor}
\begin{pf}%
The set in which we take the infimum in \eqref{ToolsEqn2} is not
empty, because it contains at least sufficiently large constants.
The remaining assertions are due to Lemma \ref{ToolsLemma1}.
\qed\end{pf}%
\begin{lem}\label{ToolsLemma3}
Let $m\in\Nset$, and let $\discr{\delta}{m}$ be the Dirac
distribution in $m$, {\ie} $\discr{\delta}{m}_m=1$ and
$\discr{\delta}{m}_l=0$ for all $l\neq{m}$. Then,
$\widehat{\discr{\delta}{m}}\in{S}_{+}$ satisfies
$\widehat{\discr{\delta}{m}}_{l} =1$ for all $l\leq{m}$. Moreover,
we have
\begin{align}
\norm{\widehat{\discr{\delta}{m}}}_{\ellInf} = 1 %
\quad\text{as well as}\quad%
\norm{\widehat{\discr{\delta}{m}}}_{\ellOne} \leq
m+\frac{\eta_0}{1-\eta_0},
\end{align}%
where $\eta_0$ is given in \eqref{ToolsEqn1a}.
\end{lem}
\begin{pf}%
At first we observe that $\widehat{\discr{\delta}{m}}$ must
decrease, so that $\discr{\delta}{m}_{l} \geq 1$ for all $l\leq{m}$.
Next we define a $S$-majorant $\sigma$ of $\discr{\delta}{m}$ by $
\sigma_l=1$ for $l\leq{m}$, and $\sigma_l=\eta_0^{\,l-m}$ for
${l}\geq{m}$. Clearly, we have $\sigma\in{S}$ as well as
\begin{align}
\norm{\sigma}_{\ellInf} = 1\quad\text{and}\quad%
\norm{\sigma}_{\ellOne} = m+\frac{\eta_0}{1-\eta_0}. 
\end{align}%
Since $\widehat{\discr{\delta}{m}}$ is the
minimal $S$-majorant of $\discr{\delta}{m}$, we conclude
$\sigma\geq\widehat{\discr{\delta}{m}}$, and this implies all
claimed results.
\qed\end{pf}%
\begin{lem}[Ball, Carr]%
\label{ToolsLemma4}%
Let $\xi\in\ellInf$ be arbitrary. Then 
$\lim\limits_{l\rightarrow\infty}{\xi_l}=0$ implies
$\lim\limits_{l\rightarrow\infty}{\widehat{\xi}_l}=0$.
\end{lem}
\begin{pf}%
Since $\widehat{\xi}$ is a nonnegative and increasing sequence there
exists the limit
\begin{align}%
2\,\eps:=\lim_{l\rra\infty}\,%
\widehat{\xi}_l=\inf_{l\in\Nset}\,\widehat{\xi}_l.
\end{align}%
Suppose for contradiction that $\eps>0$. Then there exists
$l_1\in\Nset$ such that for all $l>l_1$ we have
\begin{math}
\xi_l\leq\eps<2\,\eps\leq\widehat{\xi}_l.
\end{math}
Due to Lemma \ref{ToolsLemma3} there is at least one strictly
positive sequence $\sigma\in{S}$ with
$\lim_{l\rightarrow\infty}\sigma_l=0$. For any $b>0$ the sequence
$\eps+b\,\sigma$ is contained in $S$, and for sufficiently large $b$
it is a $S$-majorant of $\xi$, but not of $\widehat{\xi}$. This is
the desired contradiction.
\qed\end{pf}%
In the next section we need the following stronger version of Lemma
\ref{ToolsLemma4}.
\begin{lem}%
\label{ToolsLemma5}%
Any nonnegative and decreasing sequences $\xi\in\ellInf$ satisfies
\begin{align}
\label{ToolsLemma5Eqn1}%
\sum_{l=1}^{\infty}\xi_l<\infty%
\quad\Longleftrightarrow\quad%
\sum_{l=1}^{\infty}\widehat{\xi}_l<\infty.
\end{align}
\end{lem}
\begin{pf}%
$\at{\Leftarrow}$ is obvious. To prove $\at{\Rightarrow}$ let
$\xi$ be nonnegative and decreasing with $\xi\in\ellOne$. We define
a sequence $m\mapsto\discr{\sigma}{m}$ of sequences as follows
\begin{align}
\discr{\sigma}{m} := \at{\sum_{k=1}^{m-1}
\at{\xi_k-\xi_{k+1}}\,%
\widehat{\discr{\delta}{k}}}+\xi_m\,\widehat{\discr{\delta}{m}},
\end{align}
where $\widehat{\discr{\delta}{k}}$ as in Lemma \ref{ToolsLemma3}.
Clearly, for all $m$ we have $\discr{\sigma}{m}\in{S}$  and
\begin{align}
\discr{\sigma_l}{m}&=
\at{\sum_{k=1}^{m-1}
\at{\xi_k-\xi_{k+1}}\,%
\widehat{\discr{\delta}{k}}_l}+\xi_m\,\widehat{\discr{\delta}{m}}_l
\notag\\&\geq%
\at{\sum_{k=l}^{m-1}
\at{\xi_k-\xi_{k+1}}\,%
\widehat{\discr{\delta}{k}}_l}+\xi_m\,\widehat{\discr{\delta}{m}}_l
\notag\\&=\at{\sum_{k=l}^{m-1} \xi_k-\xi_{k+1}}+\xi_m=\xi_l
\end{align}
with $l=1,\,\tdots,\,m-1$. Moreover, from
$\widehat{\discr{\delta}{m+1}}\geq\discr{\delta}{m}$ it follows
$\widehat{\discr{\delta}{m+1}}\geq\widehat{\discr{\delta}{m}}$ and
hence
\begin{align}
\discr{\sigma}{m+1}=%
\discr{\sigma}{m}-\xi_{m+1}\,\widehat{\discr{\delta}{m}}+%
\xi_{m+1}\widehat{\discr{\delta}{m+1}}\geq\discr{\sigma}{m}.
\end{align}
In particular, there exists the pointwise limit
$\discr{\sigma}{\infty}:=\lim_{m\rra\infty}\discr{\sigma}{m}$. Lemma
\ref{ToolsLemma1} gives $\discr{\sigma}{\infty}\in{S}$, and with
\begin{math}
\xi\leq\discr{\sigma}{\infty}
\end{math}
we find
\begin{math}
\hat{\xi}\leq\discr{\sigma}{\infty}.
\end{math}
Finally, Lemma \ref{ToolsLemma3} provides the following uniform
estimate
\begin{align}
\norm{\discr{\sigma}{m}}_{\ellOne} &\leq%
\xi_m\,\norm{\widehat{\discr{\delta}{m}}}_{\ellOne}+
 \sum\limits_{k=1}^{m-1}
\at{\xi_k-\xi_{k+1}}\,\norm{\widehat{\discr{\delta}{k}}}_{\ellOne} %
\notag\\&\leq%
\xi_m\,\at{m+\frac{\eta_0}{1-\eta_0}}+%
\sum\limits_{k=1}^{m-1}
{\at{\xi_k-\xi_{k+1}}\,\at{k+\frac{\eta_0}{1-\eta_0}}}%
\notag\\&=%
\frac{\xi_1\,\eta_0}{1-\eta_0}+\xi_m\,m+%
\sum\limits_{k=1}^{m-1}
{\at{\xi_k-\xi_{k+1}}\,k}%
\notag\\&=%
\frac{\xi_1\,\eta_0}{1-\eta_0}+\xi_m\,m+%
\at{-\at{m-1}\,\xi_m+\sum\limits_{k=1}^{m-1}\xi_k}%
\notag\\&=%
\frac{\xi_1\,\eta_0}{1-\eta_0}+%
\sum\limits_{k=1}^{m}\xi_k\leq%
\frac{\xi_1\,\eta_0}{1-\eta_0}+\sum\limits_{k=1}^{\infty}\xi_k,
\end{align}
and the Lemma of Fatou yields $\discr{\sigma}{\infty}\in\ellOne$,
which implies $\hat{\xi}\in\ellOne$.
\qed
\end{pf}%
We mention that the equivalence \eqref{ToolsLemma5Eqn1} may fail if
$\xi$ is not decreasing.
\subsection{Proof of Theorem \ref{MassConservationTheorem1}}
Within this subsection we always assume that the assumptions of
Theorem \ref{MassConservationTheorem1} are satisfied, i.e. we have
$\lambda\at{t}\leq{R^{\,\prime}}$ for some $R^{\,\prime}<R$ and for all
$t\geq{t_0}$.
\begin{lem}
\label{MassConsLemma2}%
There exists $\mu_0<1$ and an index $l_0\in\Nset$ such that
\begin{align}
\label{MassConsEqn3} \frac{\d}{\d{t}}{\zeta}_l\at{t} &\leq
\ND{\bat{z\at{t}}}\,\gamma_{l-1}\,\frac{{q}_{l-1}}{{q}_{l}}\,%
\Bat{ \mu_0\,\bat{\zeta_{l-1}\at{t}-\zeta_l\at{t}}-
\bat{\zeta_{l}\at{t}-\zeta_{l+1}\at{t}}}
\end{align}
is satisfied for all  $l\geq{}l_0$ and all $t\geq{t_0}$.
\end{lem}
\begin{pf}%
Recall from \eqref{ASS1} that
$R=\lim_{l\rightarrow\infty}{q_l}/{q_{l+1}}$. We choose $\mu_0$ and
$l_0$ such that $\lambda\at{t}\leq\mu_0\,{q_l}/{q_{l+1}}$ holds for
all $t\geq{t_0}$ and all $l\geq{l_0}$. This implies
\begin{align}
\notag%
J_{l}\bat{z\at{t}}&=%
\GammaCondL\ND\bat{z\at{t}}\frac{{q}_{l}}{{q}_{l+1}}%
\at{%
\frac{{q}_{l+1}}{{q}_{l}}\lambda\at{t}\,
z_l\at{t}-z_{l+1}\at{t}%
}
\\&\leq{}%
\label{MassConsEqn2}%
\GammaCondL\,\ND\bat{z\at{t}}\frac{{q}_{l}}{{q}_{l+1}}%
\bat{%
\mu_0\,z_l-z_{l+1}\at{t}\at{t}%
}
\end{align} for all $t\geq{t_0}$ and $l\geq{l_0-1}$.
From Theorem \ref{ExistTheo3} we read off
\begin{align}
 \frac{\d}{\d{t}}{\zeta}_l\at{t}=J_{l-1}\bat{z\at{t}},
\end{align}
and with $z_l\at{t}=\zeta_{l}\at{t}-\zeta_{l+1}\at{t}$ we obtain
\eqref{MassConsEqn3}. %
\qed\end{pf}%
Now we can make use of the auxiliary results from the previous
subsection. For this reason we fix $\mu_0$, $t_0$ and $l_0$ as in
Lemma \ref{MassConsLemma2}, and we define the set $S$ as in Equation
\eqref{ToolsEqn1}, where the sequence $\eta$ is assumed to be
constant with value $\mu_0$. 
\par%
Our next aim is to prove that for all $\sigma\in{S_+}$ the quantity
${H}_\sigma\at{t}$ from $\eqref{MassConsEqn6}$ decreases with time
$t$.
\newcommand{\mhast}[1]{\overline{#1}}%
For any time $t$ with $t\geq{t_0}$ we define a set
$S\at{t}\subseteq{S_+}$ and a sequence $\hat{\sigma}\at{t}\in{S}$ by
\begin{align}
S\at{t}&:=\Big\{\sigma\in{S_+}\;:\;%
\sigma_l\geq\zeta_l\at{{t^{\,\prime}}}\;\;\;%
\forall\; l\geq{l_0}\;\text{and}\;\forall\;
{{t^{\,\prime}}}\in\cointerval{t}{\infty}\Big\},%
\\
\hat{\sigma}\at{t}&:=\inf\,S\at{t}.
\end{align}%
Since $S\at{t}$ contains at least the constant $\varrho_0$,
$\hat{\sigma}\at{t}\in{S}$ is well defined and satisfies
$\hat{\sigma}\at{t}\leq\varrho_0$. We mention that $t_1<t_2$ implies
$S\at{t_1}\subseteq{}S\at{t_2}$ and hence
$\hat{\sigma}\at{t_1}\geq\hat{\sigma}\at{t_2}$. \par%
For
technical reasons we introduce some discrete counterparts of $S\at{t}$
and $\hat{\sigma}\at{t}$. For fixed $m\in\Nset$ with $m\geq{l_0}$ we
define
\begin{align}%
\notag%
\discr{S}{m}\at{t}&:=\Big\{\sigma\in{S_+}\;:\;%
\sigma_l\geq\zeta_l\at{{t^{\,\prime}}}\;%
\forall\;l\;\text{with}\;{l_0}\leq{l}\leq{m+1}\;\text{and}\;
\forall\;{{t^{\,\prime}}}\in\cointerval{t}{\infty}\Big\},%
\\
\discr{\hat{\sigma}}{m}\at{t}&:=\inf\,\discr{S}{m}\at{t}.
\end{align}%
Obviously, $\discr{\hat{\sigma}}{m}\at{t}\leq\hat{\sigma}\at{t}$,
and again we find
$\discr{\hat{\sigma}}{m}\at{t_1}\geq\discr{\hat{\sigma}}{m}\at{t_2}$
for all $t_1<t_2$.  The sequence
$m\mapsto\discr{\hat{\sigma}}{m}\at{t}$ is increasing and bounded for all
$t$, because $m_1<m_2$ gives
$\hat{\sigma}\at{t}\geq\discr{\hat{\sigma}}{m_2}\at{t}%
\geq\discr{\hat{\sigma}}{m_1}\at{t}$. This implies the existence of
the pointwise limit
$\lim_{m\rra\infty}\discr{\hat{\sigma}}{m}\at{t}\leq{}\hat{\sigma}\at{t}$.
Moreover, since this limit is an $S$-majorant of
$\zeta\at{t^{\,\prime}}$ for all $t^{\,\prime}\geq{t}$, it follows
\begin{align}
\label{NewLimEqn1}
\discr{\hat{\sigma}}{m}\at{t}\xrightarrow{\;m\rra\infty\;}%
\hat{\sigma}\at{t}\quad\text{pointwise in $\ellInf$ for all $t\geq{t_0}$}.
\end{align}
\begin{rem}%
\label{NewLimRemark3}%
There exists $C\in\Rset$ such that
$\limsup_{l\rra\infty}\hat{\sigma}_l\at{t}\,l\leq{C}$ holds for all
$t\geq{t_0}$.
\end{rem}%
\begin{pf}
Let $l_1\geq{l_0}$, and define a sequence $\sigma$ by
$\sigma_l=\varrho_0$ for $l\leq{l_1}$ and  $\sigma_l=\varrho_0/l$
for $l>l_1$. For all $l>{l_1}$ we have
\begin{align}
\frac{\sigma_l-\sigma_{l+1}}{\sigma_{l-1}-\sigma_l}&=%
\frac{\D\frac{1}{l}-\frac{1}{l+1}}{\D\frac{1}{l-1}-\D\frac{1}{l}}=%
\frac{\at{l-1}}{\at{l+1}}\geq\frac{\at{l_1-1}}{\at{l_1+1}}%
\end{align}
Next we choose $l_1$ sufficiently large such that
$\at{l_1-1}/\at{l_1+1}>\mu_0$, and we  find $\sigma\in{S_+}$. Let
$t^{\,\prime}\geq{t}$ be arbitrary. By definition we have
\begin{align}
\zeta_l\at{t^{\,\prime}}=
\sum\limits_{n=l}^\infty{z}_n\at{t}%
\leq\frac{1}{l}\,\sum\limits_{n=l}^\infty%
n\,{z}_n\at{t}\leq\frac{\varrho_0}{l},
\end{align}
i.e. $\sigma$ is an $S$-majorant of $\zeta\at{t^{\,\prime}}$. Hence
$\hat{\sigma}\at{t}\leq\sigma$. Finally, $C:=\varrho_0\,l_1$
completes the proof.
\qed%
\end{pf}

\begin{rem}%
\label{NewLimRemark2}%
For fixed $m\geq{l_0}$, all $t\geq{}t_0$ and
arbitrary $\sigma\in{S_+}$ let
\begin{align}
\label{NewLimRemark2Eqn0} %
\discr{H}{m}_\sigma\at{t}:=
\max\Big\{%
\frac{\discr{\hat{\sigma}}{m}_{l_0}\at{t}}{\sigma_{l_0}},\;\,%
\frac{\discr{\hat{\sigma}}{m}_{m+1}\at{t}}{\sigma_{m+1}},\;\;%
\max\limits_{l=l_0+1,\,\tdots,\,{m}}%
\frac{\zeta_l\at{t}}{\sigma_l}\Big\}.
\end{align}
Then,
\begin{align}%
\label{NewLimRemark2Eqn1}%
\sigma\in{\discr{S}{m}}\at{t}\qquad\Longleftrightarrow\qquad
\discr{H}{m}_\sigma\at{{{t^{\,\prime}}}}\leq1\;\;\;%
\forall\;{{t^{\,\prime}}}\in\cointerval{t}{\infty}
\end{align}
is satisfied for all $\sigma\in{S_+}$.
\end{rem}
\begin{pf} %
Within this proof let ${{t^{\,\prime}}}$ always be arbitrary in
$\cointerval{t}{\infty}$. We start with $\at{\Leftarrow}$. From
$\discr{H}{m}_\sigma\at{{t^{\,\prime}}}\leq{1}$ it follows
\begin{align}
\label{NewLimRemark2Eqn2}%
\sigma_l\geq{}\zeta_l\at{{t^{\,\prime}}}%
\quad\quad\forall\;l=l_0+1,\,\tdots,\,{m},
\end{align} %
and
\begin{align}
\label{NewLimRemark2Eqn3}%
\sigma_{l_0}\geq{}%
\discr{\hat{\sigma}}{m}_{l_0}\at{{t^{\,\prime}}},\quad
\sigma_{m+1}\geq{}\discr{\hat{\sigma}}{m}_{m+1}\at{{t^{\,\prime}}}.
\end{align}%
Since
$\discr{\hat{\sigma}}{m}_{l_0}\at{{{t^{\,\prime}}}}%
\geq{}\zeta_{l_0}\at{{{t^{\,\prime}}}}$
and
$\discr{\hat{\sigma}_{m+1}}{m}\at{{{t^{\,\prime}}}}%
\geq{}\zeta_{m+1}\at{{{t^{\,\prime}}}}$ holds by construction, and
since ${{t^{\,\prime}}}$ was arbitrary, we find
$\sigma\in{\discr{S}{m}}\at{t}$. Next we prove $\at{\Rightarrow}$.
$\sigma\in{\discr{S}{m}}\at{t}$ gives
$\sigma\in{\discr{S}{m}}\at{{t^{\,\prime}}}$, and thus
$\sigma\geq\discr{\hat{\sigma}}{m}\at{{t^{\,\prime}}}$. This implies
\eqref{NewLimRemark2Eqn2} as well as \eqref{NewLimRemark2Eqn3}, and
we conclude $\discr{H}{m}_\sigma\at{{t^{\,\prime}}}\leq{1}$.
\qed%
\end{pf}
\begin{lem}
\label{NewLimLemma4}%
Let $m\geq{l_0}$ and $\sigma\in{S_+}$ be given. Then, the function
$t\mapsto\discr{H}{m}_\sigma\at{t}$ from \eqref{NewLimRemark2Eqn0}
is decreasing. In particular, any $\sigma\in{S_+}$ satisfies
\begin{align}
\label{NewLimLemma4Eqn7}%
\discr{\hat{\sigma}}{m}\at{t}\leq\discr{H}{m}_\sigma\at{t}\,\sigma.
\end{align}
\end{lem}
\begin{pf}%
Note that the function $t\mapsto\discr{H}{m}_{\sigma}\at{t}$ is well
defined and continuous for all $t\geq{t_0}$, and let $t_1\geq{t_0}$
be fixed. We prove by contradiction that
\begin{align}%
\label{MassConsEqn8}%
\discr{H}{m}_{\sigma}\at{t}<\discr{H}{m}_{\sigma}\at{t_1}+\eps
\end{align}%
holds for all $\eps>0$ and all $t\geq{}t_1$. Let $\eps>0$ be fixed
and suppose
\begin{align}%
\label{MassConsEqn5}
\discr{H}{m}_{\sigma}\at{t_2}=%
\discr{H}{m}_{\sigma}\at{t_1}+\eps=:\discr{H}{m}_\eps.
\end{align}%
for some $t_2>t_1$ with
$\discr{H}{m}_{\sigma}\at{t}<\discr{H}{m}_\eps$ for all $t$ with
$t_1\leq{t}<t_2$. We find
\begin{align}%
\label{MassConsEqn5a}
\discr{H}{m}_\eps=\max\limits_{l\geq{l_0+1},\,\tdots,\,m}%
\frac{\zeta_{l}\at{t_2}}{\sigma_l},
\end{align}
because the functions $t\mapsto\discr{\hat{\sigma}}{m}_{l_0}\at{t}$
and $t\mapsto\discr{\hat{\sigma}}{m+1}_{l_0}\at{t}$ are decreasing.
Thus there exists $l_1\in\{l_0+1,\,\tdots,\,m\}$ such that
\begin{align}%
\label{NewLimLemma4Eqn1}%
{{\zeta}_{l_1}}\at{t_2}=\discr{H}{m}_\eps\,\sigma_{l_1}.
\end{align}%
Moreover, Definition \eqref{NewLimRemark2Eqn0} guarantees that
\begin{align}%
\label{NewLimLemma4Eqn2}%
{{\zeta}_{l}}\at{t}\leq{}%
\discr{H}{m}\at{t}\,\sigma_{l}\leq{}\discr{H}{m}_\eps\,\sigma_{l}
\end{align}%
for all $l=l_0,\,\tdots,\,{m+1}$ and all
$t\in\ccinterval{t_1}{t_2}$. According to \eqref{MassConsEqn3} we
find
\begin{align} %
\frac{\d}{\d{t}}{\zeta}_{l}\at{t} %
&\leq\notag%
\ND{\bat{z\at{t}}}\,\GammaCondL\,\frac{{q}_{l}}{{q}_{l+1}}\, \Bat{
\mu_0\,\zeta_{l-1}\at{t}-\at{1+\mu_0}\,
\zeta_{l}\at{t}+\zeta_{l_1+1}\at{t}}
\\&\leq\notag%
\ND{\bat{z\at{t}}}\,\GammaCondL\,\frac{{q}_{l}}{{q}_{l+1}}\, \Bat{
\discr{H}{m}_\eps\,\bat{\mu_0\,\sigma_{l-1}+\sigma_{l+1}}-\at{1+\mu_0}\,
\zeta_l\at{t}}
\\&\leq\label{NewLimLemma4Eqn5}%
\varrho_0\,\GammaCondL\,\frac{{q}_{l}}{{q}_{l+1}}\,\at{1+\mu_0}
\Bat{ \discr{H}{m}_\eps\,\sigma_l-\zeta_l\at{t}},
\end{align}
where the last estimate is due to
$\mu_0\,\sigma_{l-1}+\sigma_{l+1}\leq\at{1+\mu_0}\,\sigma_{l_1}$
which follows from the definition of $S$, see \eqref{ToolsEqn1}. We
apply Gronwall's Lemma for $t\in\ccinterval{t_1}{t_2}$, and obtain
\begin{align}
\label{NewLimLemma4Eqn3}%
\Bat{\discr{H}{m}_\eps\,%
\sigma_l-\zeta_l\at{t_2}}\geq\exp\bat{-c_l\,\at{t_2-t_1}}\,
\Bat{\discr{H}{m}_\eps\,%
\sigma_l-\zeta_l\at{t_1}}>0,
\end{align}
where $c_l>0$ can be read off from \eqref{NewLimLemma4Eqn5}. The
estimate \eqref{NewLimLemma4Eqn3} with $l=l_1$ is a contradiction
for
\eqref{NewLimLemma4Eqn1}. Thus we have proved \eqref{MassConsEqn8}, %
and the limit $\eps\rightarrow{0}$ yields the claimed monotonicity
result. Finally, let $\sigma\in{S_+}$ be fixed and $t\geq{t_1}$ be
arbitrary. We find
\begin{align}
\zeta_l\at{t}\leq\discr{H}{m}_\sigma\at{t}\,\sigma%
\leq\discr{H}{m}_\sigma\at{t_1}\;\sigma
\quad\text{for all $l=l_0,\,\tdots,\,m+1$}.
\end{align}
In particular,
$\discr{H}{m}_\sigma\at{t_1}\,\sigma\in\discr{S}{m}\at{t_1}$,
and it follows
$\discr{\hat{\sigma}}{m}\at{t_1}\leq%
\discr{H}{m}_\sigma\at{t_1}\,\sigma$,
which was claimed in \eqref{NewLimLemma4Eqn7}.
\qed%
\end{pf}
\begin{lem}
\label{NewLimLemma5}%
For all $t\geq{t_0}$ we have
\begin{math}
\hat{\sigma}\at{t}=\widehat{\eta\at{t}}
\end{math}
where \begin{align}
\label{NewLimLemma5Eqn1}%
\eta\at{t}:=\max\big\{\zeta\at{t},\,\hat{\sigma}_{l_0}\at{t}\,
\discr{\delta}{l_0}\big\},
\end{align}%
$\discr{\delta}{l_0}$ is the Dirac distribution in $l_0$,
and $\widehat{\eta\at{t}}$ is the
minimal $S$-majorant of $\eta\at{t}$.
\end{lem}%
\begin{pf}%
Let $m\geq{l_0}$ be arbitrary.
Remark \ref{NewLimRemark2} and Lemma \ref{NewLimLemma4} provide%
\begin{align}
\label{NewLimLemma5Eqn3}%
\discr{S}{m}\at{t}&=%
\Big\{\sigma\in{S_+}\;:\;\discr{H}{m}\at{t}\leq{1}\Big\}
\notag\\&=%
\left\{%
\sigma\in{S_+}\;:\;\begin{array}{l}
\sigma_{l_0}\geq\discr{\hat{\sigma}}{m}_{l_0}\at{t}\\
\sigma_{l}\geq\zeta_l\at{t}
\;\forall\;l=l_0+1,\,\tdots,\,m\\%
\sigma_{m+1}\geq\discr{\hat{\sigma}}{m}_{m+1}\at{t}%
\end{array} \right\}.
\end{align}
Let $\discr{\eta}{m}\at{t}\in{S_+}$ be defined by
\begin{align}
\label{NewLimLemma5Eqn4}%
\discr{\eta}{m}\at{t}:=%
\hat{\sigma}_{m+1}\at{t}+\widehat{\eta\at{t}},
\end{align}
with $\eta\at{t}$ as in \eqref{NewLimLemma5Eqn1}. As simple
calculation shows
$\discr{H}{m}_{\discr{\eta}{m}\at{t}}\at{t}\leq{1}$, and from
\eqref{NewLimLemma4Eqn7} it follows that
\begin{align}
\label{NewLimLemma5Eqn5}%
\discr{\hat{\sigma}}{m}\at{t}\leq%
\discr{H}{m}_{\discr{\eta}{m}\at{t}}\at{t}\,\discr{\eta}{m}\at{t}%
\leq\widehat{{\eta}\at{t}}+\hat{\sigma}_{m+1}\at{t}.
\end{align}
According to Remark \ref{NewLimRemark3} and \eqref{NewLimEqn1} the
limit $m\rra\infty$ provides
\begin{align}
\hat{\sigma}\at{t}=%
\lim\limits_{m\rra\infty}\discr{\hat{\sigma}}{m}\at{t}\leq
\widehat{\eta\at{t}}.
\end{align}
Moreover, by construction we have
$\hat{\sigma}_l\at{t}\geq\zeta_l\at{t}$ and
$\hat{\sigma}_l\at{t}\geq\hat{\sigma}_{l_0}\,\discr{\delta}{l_0}_l$
for all $l\geq{l_0}$, which shows that $\hat{\sigma}\at{t}$ is
$S$-majorant for $\eta\at{t}$. Therefore,
$\hat{\sigma}\at{t}\geq\widehat{\eta\at{t}}$.
\qed%
\end{pf}
\begin{cor}
\label{NewLimCor6} For all $t\geq{t_0}$ and all $\sigma\in{S_+}$ let
${H}_\sigma\at{t}$ be given as in $\eqref{MassConsEqn6}$, i.e.
\begin{align}
\label{NewLimCor6Eqn1}%
{H}_\sigma\at{t}:=
\max\left\{\frac{\varrho_0}{\sigma_{l_0}},\,%
\sup\limits_{l\geq{l_0+1}}
\frac{\zeta_{l}\at{t}}{\sigma_l}\right\},
\end{align}
where ${H}_\sigma\at{t}$ may be infinite. Then, for all
$\sigma\in{S_+}$ the function $t\mapsto{H}_\sigma\at{t}$ is
decreasing. In particular, any $\sigma\in{S_+}$ satisfies
\begin{align}
\label{NewLimCor6Eqn2}%
\hat{\sigma}\at{t}\leq{H}_\sigma\at{t}\,\sigma.
\end{align}
\end{cor}
\begin{pf}
Let $\sigma\in{S_+}$ be fixed, let $t_1\geq{t_0}$ with
$H:=H_\sigma\at{t_1}<\infty$, and let $t_2\geq{t_1}$. With
$\tilde{\sigma}=H\sigma$ we find
\begin{math}
\tilde{\sigma}_{l_0}\geq\varrho_0\geq%
\hat{\sigma}_{l_0}\at{t_1}\geq{}\zeta_{l_0}\at{t_1} %
\end{math}
and $\tilde{\sigma}_l\geq\zeta_l\at{t_1}$ for all $l>{l_0}$. In
particular, $\tilde{\sigma}$ is an $S$-majorant of $\eta\at{t_1}$,
and we conclude $\tilde{\sigma}\geq\widehat{\eta\at{t_1}}$. Lemma
\ref{NewLimLemma5} yields
$\tilde{\sigma}\geq\hat{\sigma}\at{t_1}\geq\hat{\sigma}\at{t_2}$,
and it follows that $\tilde{\sigma}$ is an $S$-majorant of $\zeta\at{t_2}$.
This implies $\tilde{\sigma}_l\geq\zeta_l\at{t_2}$ for all
$l>{l_0}$, and hence
\begin{align}
1\geq{}H_{\tilde\sigma}\at{t_2}=\frac{1}{H}\,H_{\sigma}\at{t_2},
\end{align}
which was claimed. Finally, \eqref{NewLimCor6Eqn2} follows
immediately.
\qed%
\end{pf}
\begin{cor} Let $\discr{z}{\infty}$ be as in Corollary \ref{LimCorr2}, and let $\eps>0$ be arbitrary.
Then we have
\label{NewLimCorr8}%
\begin{align}
\label{NewLimCorr8Eqn1}%
\varrho_0\geq\varrho\at{\discr{z}{\infty}}\geq\varrho_0-2\eps
\end{align}
In particular, Theorem \ref{MassConservationTheorem1} is proved.
\end{cor}
\begin{pf}
Let $\eps>0$ be fixed, and let
\begin{align}
\sigma:=H\,\widehat{\zeta}\at{t_0},\quad
H:=H_{\widehat{\zeta\at{t_0}}}\at{t_0}=%
\frac{\varrho_0}{\widehat{\zeta\at{t_0}}_{l_0}}.
\end{align}
We have $H\in\cointerval{1}{\infty}$, and Lemma \ref{ToolsLemma5}
provides $\sigma\in\ellOne$. Therefore we can choose an index
$l_1\geq{l_0}$ such that $\sum_{l=l_1}^\infty\sigma_l\leq{\eps}$.
Moreover, according to Corollary \ref{NewLimCor6} we have 
$\zeta_l\at{t}\leq\sigma_l$ for all $l\geq{l_0}$ and all $t\geq{t_0}$. Therefore, we find
\begin{align}
\varrho_0&=\sum\limits_{l=1}^{\infty}\,\zeta_l\at{t}\leq
\eps+\sum\limits_{l=1}^{l_1}\,\zeta_l\at{t}=
\eps+\sum\limits_{l=1}^{l_1}\,\NL_l\bat{z\at{t}}.
\end{align}
By construction, there exists a sequence $m\mapsto{t_m}$ with $t_m\rra\infty$ such that
$z\at{t_m}\rra\discr{z}{\infty}$ weak$^\star$ in $X$ for $m\rra\infty$. Using the weak$^\star$
continuity of the functionals $\NL_l$ we find
\begin{align}
\sum\limits_{l=1}^{l_1}\,\NL_l\at{\discr{z}{\infty}}\geq\varrho_0-\eps.
\end{align}
Finally, it is easy to prove that $\varrho\at{\discr{z}{\infty}}\leq\varrho_0$ implies
\begin{math}
\varrho\at{\discr{z}{\infty}}=
\sum_{l=1}^{\infty}\NL_l\at{\discr{z}{\infty}},
\end{math}
and \eqref{NewLimCorr8Eqn1} follows immediately. 
\qed\end{pf}%

%% file: HNN_2005.bbl
\begin{thebibliography}{10}
\expandafter\ifx\csname url\endcsname\relax
  \def\url#1{\texttt{#1}}\fi
\expandafter\ifx\csname urlprefix\endcsname\relax\def\urlprefix{URL }\fi

\bibitem{BD35}
R.~Becker, W.~D\"oring, {Kinetische Behandlung der Keimbildung in
  \"ubers\"attigten D\"ampfen}, Ann. Phys. (Leipzig) 4 (1935) 719--752.

\bibitem{Fre39}
J.~I. Frenkel, A general theory of heterophase fluctuations and pretransition
  phenomena, Journal of Chemical Physics 7 (1939) 538--547.

\bibitem{DD05}
W.~Dreyer, F.~Duderstadt, On the {Becker}/{D\"oring} theory of nucleation of
  liquid droplets in solids, to appear in J. Stat. Phys (2005).

\bibitem{Bur77}
J.~J. Burton, Nucleation theory, in: Statistical Mechanics, {Part A}:
  Equilibrium techniques, Plenum Press, New York, London, 1977, pp. 195--234.

\bibitem{PL79}
O.~Penrose, J.~Lebowitz, Towards a rigorous theory of metastability, in:
  Studies in statistical mechanics, {Vol. VII}: Fluctuation phenomena,
  North--Holland, Amsterdam, 1979, pp. 293--340.

\bibitem{Pen89}
O.~Penrose, Metastable states for the {Becker}--{D\"oring} cluster equations,
  Comm. Math. Phys. 124 (1989) 515--541.

\bibitem{CDW95}
J.~Carr, D.~B. Duncan, C.~H. Walshaw, Numerical approximation of a metastable
  system, IMA J. Numer. Anal. 15~(4) (1995) 505--521.

\bibitem{Sle00}
M.~Slemrod, The {Becker}-{D\"oring} equations, in: N.~Bellomo, M.~Pulvirenti
  (Eds.), Modeling in applied sciences, Birkh\"auser, Boston, 2000, pp.
  149--171.

\bibitem{Nie03}
B.~Niethammer, On the evolution of large clusters in the {Becker}-{D\"oring}
  model, J. Nonlinear Science 13~(1) (2003) 115--155.

\bibitem{BCP86}
J.~Ball, J.~Carr, O.~Penrose, The {Becker}-{D\"oring} cluster equations: Basic
  properties and asymptotic behaviour of solutions, Commun. Math. Phys. 104
  (1986) 657--692.

\bibitem{Nal05}
M.~Naldzhieva, {Die thermo\-dyna\-misch konsist\-enten Becker--D\"oring
  Gleichungen}, Diploma thesis, Humboldt-Universit\"at zu Berlin, Department of
  Mathematics, in preparation (2005).

\bibitem{LM02}
P.~Lauren\c{c}ot, S.~Mischler, From the {Becker}--{D\"oring} to the
  {Lifshitz}--{Slyozov}--{Wagner} equations, J. Stat. Phys 106 (2002) 957--991.

\bibitem{Pen97}
O.~Penrose, The {Becker}--{D\"oring} equations at large times and their
  connection with the {LSW} theory of coarsening, J. Stat. Phys. 89~(1/2)
  (1997) 305--320.

\bibitem{CGPV02}
J.-F. Collet, T.~Goudon, F.~Poupaud, A.~Vasseur, The {Becker}--{D\"oring}
  system and its {Lifshitz}--{Slyozov} limit, SIAM J. Appl. Math 62~(5) (2002)
  488--1500.

\bibitem{Nie04}
B.~Niethammer, A scaling limit of the {Becker}-{D\"oring} equations in the
  regime of small excess density, J. Nonlinear Science 14~(5) (2004) 453--468.

\bibitem{BC88}
J.~Ball, J.~Carr, Asymptotic behavior of solutions to the {Becker}--{D\"oring}
  equations for arbitrary initial data, Proc. R. Soc. Edinb., Sect. A 108~(1/2)
  (1988) 109--116.

\bibitem{Can05}
J.~A. {Canizo~Rinc\'on}, Asymptotic behavior of solutions to the generalized
  {Becker-D\"oring} equations for general initial data, to appear in Proc. R.
  Soc. Edinb., Sect. A (2005).
\newline\urlprefix\url{"www.hyke.org/preprint/2005/08/080.pdf"}

\end{thebibliography}
